%% file: main.tex
\documentclass[UTF8,a4paper,onecolumn,11pt,accepted=2021-08-28]{quantumarticle}
\pdfoutput=1
\usepackage[utf8]{inputenc}
\usepackage[english]{babel}
\usepackage[T1]{fontenc}
\usepackage{amsmath}
\usepackage{amsthm}
\usepackage{amssymb}
\usepackage{booktabs}
\usepackage{braket}
\usepackage[pagebackref,bookmarks=true]{hyperref}
\usepackage[linesnumbered,lined,ruled,commentsnumbered]{algorithm2e}
\usepackage[capitalise]{cleveref}
\usepackage{comment}
\usepackage{complexity}
\usepackage[shortlabels]{enumitem}
\usepackage{footnotebackref}
\usepackage{mathtools}
\usepackage{multirow}
\usepackage{longtable}
\usepackage{subcaption}
\usepackage{tocbibind}
\usepackage[numbers,compress]{natbib}

\usepackage{tikz}
\usepackage{lipsum}

\usepackage[normalem]{ulem}

\usetikzlibrary{decorations.pathreplacing,matrix,tikzmark}

\theoremstyle{definition}
\newtheorem{example}{Example}[section]
\newtheorem{definition}{Definition}[section]

\DeclarePairedDelimiter{\ceil}{\lceil}{\rceil}

\input{macros}

\tikzstyle{decision} = [diamond, draw, fill=blue!20, 
    text width=4.5em, text badly centered, node distance=3cm, inner sep=0pt]
\tikzstyle{block} = [rectangle, draw, fill=blue!20, 
    text width=5em, text centered, rounded corners, minimum height=4em]
\tikzstyle{line} = [draw, -latex']
\tikzstyle{cloud} = [draw, ellipse,fill=red!20, node distance=3cm,
    minimum height=2em]

\SetKwInput{KwData}{Input}
\SetKwInput{KwResult}{Output}
\SetArgSty{textup}
\DontPrintSemicolon

\begin{document}

\title{Quantum-accelerated constraint programming}

\author{Kyle E. C. Booth}
\affiliation{Quantum Artificial Intelligence Laboratory (QuAIL), NASA Ames Research Center, Moffett Field, CA 94035, USA}
\affiliation{USRA Research Institute for Advanced Computer Science (RIACS), Mountain View, CA 94043, USA}
\email{kyle.booth@nasa.gov}
\orcid{}
\author{Bryan O'Gorman}
\affiliation{Quantum Artificial Intelligence Laboratory (QuAIL), NASA Ames Research Center, Moffett Field, CA 94035, USA}
\affiliation{Berkeley Quantum Information and Computation Center, University of California, Berkeley, CA 94720, USA}
\email{bogorman@berkeley.edu}
\author{Jeffrey Marshall}
\affiliation{Quantum Artificial Intelligence Laboratory (QuAIL), NASA Ames Research Center, Moffett Field, CA 94035, USA}
\affiliation{USRA Research Institute for Advanced Computer Science (RIACS), Mountain View, CA 94043, USA}
\email{jeffrey.s.marshall@nasa.gov}
\author{Stuart Hadfield}
\affiliation{Quantum Artificial Intelligence Laboratory (QuAIL), NASA Ames Research Center, Moffett Field, CA 94035, USA}
\affiliation{USRA Research Institute for Advanced Computer Science (RIACS), Mountain View, CA 94043, USA}
\email{stuart.hadfield@nasa.gov}
\author{Eleanor Rieffel}
\affiliation{Quantum Artificial Intelligence Laboratory (QuAIL), NASA Ames Research Center, Moffett Field, CA 94035, USA}
\email{eleanor.rieffel@nasa.gov}

\maketitle
\input{sections/0-abstract.tex}

\input{sections/1-introduction.tex}
\input{sections/2-background.tex}

\input{sections/3-related-work.tex}
\input{sections/4-quantum-access.tex}
\input{sections/5-filtering.tex}

\input{sections/6-backtracking-search.tex}

\input{sections/7-conclusions.tex}
\input{sections/8-acknowledgements.tex}

\bibliographystyle{unsrtnat}
\bibliography{root}

\appendix
\input{sections/appendix/quantum_tarjan.tex}

\input{sections/appendix/other_globals.tex}
\input{sections/appendix/models.tex}

\input{sections/appendix/unitarizing_quantum_filtering}

\end{document}

%% file: macros.tex
\DeclareMathOperator*{\argmin}{arg\,min}

\newcommand{\dfsindex}{\textsc{index}}
\newcommand{\lowlink}{\textsc{ll}}
\newcommand{\ngbrs}{N}
\newcommand{\degree}{\delta}

\newcommand{\indeterminate}{*}
\newcommand{\nullassign}{*}

\newcommand{\depth}{L}

\newcommand{\outcome}{{e}}
\newcommand{\outcomes}{{\mathbf e}}
\newcommand{\outcomeSet}{{E}}
\newcommand{\outcomeSets}{{\mathcal E}}

\newcommand{\diffusion}{W}
\newcommand{\walk}{W}

\newcommand{\flag}{\beta}

\newcommand{\Dorn}{D\"{o}rn}
\newcommand{\Durr}{D\"{u}rr}
\newcommand{\Hoyer}{H{\o}yer}
\newcommand{\Regin}{R\'{e}gin}

\newcommand{\heuristic}{h}
\newcommand{\numChildren}{\heuristic_{\mathrm{nu}}}
\newcommand{\numEldestChildren}{\tilde{\heuristic}_{\mathrm{nu}}}
\newcommand{\propagate}{F}
\newcommand{\predicate}{\propagate_{\mathrm{p}}}
\newcommand{\filter}{\propagate_{\mathrm{d}}}

\newcommand{\uniform}{\mathrm{unif}}
\newcommand{\branch}{b}
\newcommand{\maxNumChildren}{B}
\newcommand{\child}{\mathrm{ch}}

%% file: sections/0-abstract.tex
\begin{abstract}
Constraint programming (CP) is a paradigm used to model and solve constraint satisfaction and combinatorial optimization problems. 
In CP, problems are modeled with constraints that describe acceptable solutions and solved with backtracking tree search augmented with logical inference. 
In this paper, we show how quantum algorithms can accelerate CP, at both the levels of inference and search.
Leveraging existing quantum algorithms, we introduce a quantum-accelerated filtering algorithm for the \texttt{alldifferent} global constraint and discuss its applicability to a broader family of global constraints with similar structure.
We propose frameworks for the integration of quantum filtering algorithms within both classical and quantum backtracking search schemes, including a novel hybrid classical-quantum backtracking search method. 
This work suggests that CP is a promising candidate application for early fault-tolerant quantum computers and beyond. 
\end{abstract}

%% file: sections/1-introduction.tex
\section{Introduction}

Constraint programming (CP) is a paradigm used to model and solve constraint satisfaction and combinatorial optimization problems~\cite{rossi2006handbook}. In CP, a problem is expressed in terms of a declarative model, identifying variables and constraints, and the model is evaluated using a general-purpose constraint solver, leveraging techniques from a variety of fields including artificial intelligence, computer science, and operations research.
CP has successfully been used to approach challenging problems in areas such as scheduling, planning, and vehicle routing~\cite{rossi2006handbook,baptiste2012constraint,laborie2018ibm}.

Given a CP model of a problem, a constraint solver performs a search for values of the variables that satisfy the expressed constraints. The search performed is often systematic, as is the case in a backtracking search~\cite{van2006backtracking}, although some solvers employ incomplete schemes such as local search~\cite{hentenryck2009constraint,bjordal2015constraint}. In this work, we focus on the former, with an emphasis on backtracking search. While CP bears similarity to other paradigms for modeling and solving combinatorial problems, such as boolean satisfiability (SAT)~\cite{biere2009handbook} or integer programming (IP)~\cite{wolsey1998integer}, the technology differentiates itself in a number of key ways. Modeling efforts leverage rich decision variable types and focus on identifying and combining encapsulations of frequently occurring combinatorial substructure. 
Search effort is reduced through logical inference, a process whereby possible variable-value assignments are ruled out based on the constraints in the model. 
Each constraint is supported by an inference algorithm that rules out value assignments for the variables in the scope of the constraint. Assignments removed in one constraint often propagate to other constraints, allowing even more inference to be performed~\cite{rossi2006handbook}. This emphasis on inference and propagation is particularly powerful as it can significantly reduce the fraction of the search space that must be explored, detecting ``dead ends'' as early as possible. 

It follows that the success of CP is dependent on the availability of efficient procedures for performing inference to prune infeasible value assignments. 
The benefits of a reduced search space are severely muted if the time taken to perform inference exceeds the time needed to explore the area pruned by that inference.
Indeed, a large body of research exists that is expressly focused on finding increasingly efficient inference algorithms for various constraints~\cite{regin1994filtering,cymer2012dulmage,quimper2004improved,beldiceanu2007global}.
In this paper, we explore the use of quantum computing to accelerate CP, focusing on innovations for both inference and search. 

In the context of inference, we explore the use of quantum algorithms for graph problems, especially that for finding maximum matchings in graphs~\cite{dorn2009quantum}, to accelerate classical inference algorithms in CP.\@ The quantum algorithms used heavily exploit Grover's algorithm for unstructured search~\cite{grover1996fast}. In addition to speeding up inference, we argue that the structure of inference in the CP paradigm represents an attractive framework for the deployment of quantum algorithms. The encapsulation of combinatorial substructure within CP models provides an elegant mechanism for carving off portions of complex problems into inference subproblems that can be solved by a quantum co-processor. 
These smaller subproblems require fewer resources, making them promising candidates for the early fault-tolerant quantum computers of the future.
With respect to search, we investigate the adaptation of existing quantum tree search algorithms~\cite{montanaro2015quantum,jarret2018improved} to the search performed within CP, and provide preliminary resource estimates for this integration. Our adaptations are focused on incorporating quantum filtering within both classical and quantum backtracking search algorithms.

At a high level, this paper is intended to provide researchers in quantum computing with an introduction to core concepts in CP before detailing how quantum algorithms can be applied to the paradigm. 
While CP has been recently used as an approach to more efficiently compile quantum circuits~\cite{booth2018comparing}, this work, along with an earlier version~\cite{booth2020quantum}, conducts the first investigation into the formal integration of quantum computing into CP.\@ 
Our initial explorations indicate the potential for symbiosis between the two paradigms: quantum algorithms can accelerate both inference and search in CP, and CP offers an attractive, modular formalism for tackling hard problems that makes it a promising candidate application for early fault-tolerant quantum computers,\footnote{By ``early fault-tolerant'' quantum computers, we mean future devices with error rates small enough, for example, to enable phase estimation, but with a limited number of logical qubits. While we argue that our proposals are suitable for early generations of such devices because their hybrid nature allows for putting smaller parts of the problem on the device, we do not expect that the quantum algorithms we discuss will be successfully implementable using NISQ devices. 
In other words, our target is quantum devices that are still intermediate-scale but for which noise (at the logical level) is not a significant factor.} and beyond.   

This paper primarily combines and exploits existing quantum algorithms from the literature, and some extensions thereto, to provide quantum algorithms for CP.
Our main overall contribution is a detailed examination of how these algorithmic building blocks can be put together and applied to CP. This includes a careful analysis of subtle aspects of this application, such as parameter regimes in which the algorithms best classical algorithms, the use of classical data in quantum algorithms, and how to incorporate filtering into backtracking when one or both are quantum. While the details of this analysis are specific to this application, we expect that our approach to the subtler aspects of these algorithms will be useful for designing and analysing the use of these building blocks in other applications.

The primary contributions of this paper are as follows:

\begin{enumerate}[i.]
    \item 
        We propose a quantum-accelerated $\tilde{O}(\sqrt{|X||V||E|})$-time\footnote{We use the notation $\tilde{O}(f(n)) = O(f(n) \polylog f(n))$ to suppress factors that are polylogarithmic in the dominant part of the scaling.} bounded-error algorithm for domain consistency of the \texttt{alldifferent} constraint, where $|X|$ is the number of variables, $|V|$ is the number of unique domain values, and $|E|$ is the sum of the variables' domain sizes.
Our approach follows the main classical algorithm, accelerating the basic subroutines performed at each iteration with quantum analogs.
The complexity is dominated by that for finding maximum matchings in bipartite graphs.
Long-standing state-of-the-art deterministic and randomized classical algorithms take $O(\sqrt{|X|}|E|)$ and $O({|X|}^{\omega-1} |V|)$ time, respectively, 
where $\omega$ corresponds to the asymptotic cost of classical matrix multiplication; the best upper bound known on $\omega$ is $2.373$.\footnote{We note that the instance size at which the asymptotic scaling becomes relevant is so large that in, practice, the cost of matrix multiplication may scale cubically.}  Our approach, leveraging an existing quantum algorithm, improves over these 
bounds by factors on the order of
$\sqrt{|E|/|V|}$
and
$\sqrt{|X|^{2\omega-3}|V|/|E|}$,
respectively,
up to polylogarithmic terms.\footnote{For the purpose of algorithm comparison we consider the ratio of worst-case time-complexity upper bounds. Specifically, an algorithm with time-complexity bound $\tilde{O}(g(n))$ is said to improve upon an $\tilde{O}(f(n))$ algorithm by a factor of $f(n)/g(n)$.} 
A recently proposed classical interior-point-based method (IPM), improving on the aforementioned state-of-the-art, takes $\tilde{O}(|E| + |V|^{3/2})$ time. For the regime where $|E|=O(|V|^{3/2})$, our approach improves over this method by a factor on the order of $|V|/\sqrt{|X||E|}$, up to polylogarithmic terms. Within this regime, when $|X|=O(\sqrt{|V|})$ our approach always improves over IPM, while $|X|=\Omega(\sqrt{|V|})$ yields an improvement when $|X||E|=O(|V|^2)$. 
\item 
We identify a broader family of global constraints, including the global cardinality constraint (\texttt{gcc}) and the \texttt{inverse} constraint, whose domain consistency filtering algorithms can be accelerated with quantum algorithms. As with the \texttt{alldifferent} constraint, the worst-case complexity of the classical domain consistency filtering algorithms for these global constraints is dominated by finding maximum matchings in bipartite graphs. 
\item 
We detail frameworks for integrating quantum-accelerated inference algorithms within classical and quantum backtracking search schemes. We show that the speedups noted for previously proposed quantum backtracking algorithms can also be leveraged for quantum branch-and-infer search. We also propose partially quantum search schemes that yield speedups for smaller sections of the tree, intended for early, resource-constrained quantum devices. Finally, we provide preliminary resource estimates and discuss the benefits and drawbacks of each search approach.  
\end{enumerate}

The organization of the paper is as follows. Section~\ref{sec:background} provides background for important concepts in CP and Section~\ref{sec:related-work} summarizes relevant previous work. 
Section~\ref{sec:qram} discusses how quantum algorithms can access classical data. 
Section~\ref{sec:filtering} details a quantum-accelerated algorithm for the \texttt{alldifferent} constraint and discusses its generalization to other global constraints with a similar structure. Section~\ref{sec:tree-search} details the integration of our quantum-accelerated filtering algorithms within both classical and quantum tree search schemes. Finally, Section~\ref{sec:conclusions} provides concluding remarks and future research directions. 

%% file: sections/2-background.tex
\section{Constraint Programming Background}\label{sec:background}

In this section we provide background on the fundamental concepts of constraint programming (CP). 
The interested reader is referred to additional sources for a more thorough review of the subject~\cite{rossi2006handbook,van2006global}. 
There is a variety of open-source~\cite{perron2011operations,nethercote2007minizinc,gecode2019gecode,chu2019chuffed} and commercial~\cite{laborie2018ibm} software available for modeling and solving problems using CP. 

\subsection{Constraint satisfaction problems}\label{sec:csp}

CP is a paradigm used for solving constraint satisfaction and optimization problems. A constraint satisfaction problem (CSP) consists of variables $X = (x_1,\ldots,x_{|X|})$, with associated domains $\mathcal{D} =(D_1,\ldots,D_{|X|})$, and constraints $\mathcal{C}= (C_1,\ldots,C_{|\mathcal{C}|})$. 
The domain $D_i = \{d_{i, 1}, \dots, d_{i, |D_i|}\}$ of variable $x_i$ is the finite set of values the variable can possibly be assigned. 
Each constraint $C \in \mathcal{C}$ acts on a subset of the variables $X$, known as the scope of the constraint. 
Let $d_i^*$ represent the value actually assigned to variable $x_i$.
A solution to a CSP is a tuple of assigned values $(d_1^*,\ldots,{d}_{|X|}^*) \in D_1 \times \cdots \times D_{|X|}$ such that, for each constraint $C \in \mathcal{C}$ the values assigned to the variables within its scope satisfy the constraint.

Similarly, a constraint optimization problem (COP) is a CSP with an objective function. The solution to a COP is an assignment of values to variables, from the domains of those variables, such that each constraint is satisfied and the objective function is optimized (either maximized or minimized). The objective function is typically represented by a variable and associated domain. In this work we focus on techniques that can be employed to solve CSPs and COPs within a backtracking tree search framework.

\subsection{Backtracking search algorithms}\label{sec:backtracking}

Backtracking search algorithms are an important and general class of algorithms used to solve CSPs and can be seen as performing a depth-first traversal of a search tree~\cite{van2006backtracking}. The search tree provides a systematic way of investigating different decisions in a divide-and-conquer fashion. The root node of the search tree corresponds to the original CSP and subsequent nodes, generated via \emph{branching}, are increasingly constrained versions of the original CSP. Backtracking algorithms are effective given  the means to quickly test whether or not a node in the search tree can lead to a solution; for this reason, backtracking search algorithms are not helpful for unstructured search problems. A node that cannot lead to a solution is called a \emph{dead end}. In general, it is advantageous to detect dead ends as quickly as possible so that search effort can be directed to more promising areas of the search tree.

In the simplest form of backtracking, often called naive backtracking \cite{van2006backtracking}, a branch represents an assignment of a value to an unassigned variable. 
This can be thought of as adding a constraint to the CSP, i.e., $x_i = d_j$ for some $d_j \in D_i$, or, alternatively, as reducing the domain associated with the variable to the assigned value, i.e., $D_i = \{d_j\}$. 
There are also more sophisticated branching rules than this simple assignment-based branching.
For each non-leaf node in the tree, a child is generated for each value in the domain of the variable being branched on. 
The branching process involves variable- and value-selection heuristics; the former identifies the variable to branch on, and the latter identifies the order in which the branches are explored.

The node resulting from the branch corresponds to the CSP of the parent with an updated domain. A predicate $P$ is then used to test whether the node can lead to a solution or not, returning $1$ if the node indicates a solution to the CSP, $0$ if the node cannot lead to a solution (i.e., definitely infeasible), and indeterminate ($\indeterminate$) otherwise. The ability to efficiently determine the value of $P$ is often due to exploiting problem structure.

If a node in the search is a dead end (i.e., $P$ returns $0$), the most recent branch is undone (backtracked), a new branch is posted, and the process repeats. If the node represents a solution to the CSP (i.e., $P$ returns $1$), the problem has been solved.\footnote{In the case of a COP, the objective value of the solution is recorded and backtracking continues in search of a better solution (or proof that none exists).} 
If it is not clear that the node can lead to a solution (i.e., $P$ returns $\indeterminate$), a new branch is posted, the resulting node is tested with $P$, and the process repeats.

Thus, there are two core ingredients of backtracking search: i) a means of testing whether a node can lead to a solution or not, and ii) heuristics that determine how to branch.  In the next section we detail the branch-and-infer tree search used in CP.

\subsection{Branch-and-infer search}\label{sec:background-branch-infer-search}

Search effort in CP is reduced via logical inference.
Each constraint in the CSP model is supported by an inference algorithm that performs domain \emph{filtering} (also known as \emph{pruning}), a process that removes possible variable-value assignments by proving that they cannot satisfy the constraint, and thus cannot participate in a solution to the CSP.
As variables often participate in multiple constraints, value removal by the filtering of one constraint often triggers value removals from variables in neighboring constraints in a process termed \emph{propagation}. For the purposes of this paper, the term logical inference encapsulates both filtering and propagation. The remainder of this section provides a more formal description of the overall search process.

CP's branch-and-infer search follows the framework of backtracking search; however, the predicate $P$ is extended to a \emph{propagation function} $F$ that prunes values from the domains of the variables and, through this domain reduction, determines whether a node can lead to a solution or not. The search is specified by two main operators: the propagation function $F$ and a branching operator $h$. 

Let us define 
\begin{equation}
2^{\mathcal D} :=
\times_{i=1}^{n} 2^{D_i}
\end{equation}
where $2^S$ is the power set of set $S$, i.e., $2^{\mathcal D}$ is the set of tuples of subsets of the domains. The search proceeds by exploring a tree in which each node is associated with the variables and constraints of the original CSP, and a distinct element of $2^{\mathcal D}$ representing the domains ``active'' at that node.

\subsubsection{Propagation Function} The propagation function
\begin{equation}\label{eq:F}
F: 2^{\mathcal D} \to
2^{\mathcal D} \times \{0,1,\indeterminate\}
\end{equation}
takes active domains\footnote{
The set $2^\mathcal{D}$ denotes the set of global domain tuples throughout the paper. For simplicity of presentation, we overload the symbol $\mathcal{D}$ to denote active domains (i.e., elements of $2^\mathcal{D}$).}
 $\mathcal{D}$ and returns
filtered domains $\mathcal D'$ and a flag $\flag \in \{0, 1, \indeterminate\}$ representing the status of the node (definitely infeasible, definitely feasible, or indeterminate, respectively). We also use $ \filter(\mathcal D)$ and $\predicate(\mathcal D)$ to denote the filtered domains and flag (predicate value) returned by $F$, respectively. A description of the propagation function $F$ is given by Algorithm \ref{alg:filter-domains}. 

The propagation function first initializes the node status to indeterminate ($\indeterminate)$. Then, it enters a loop of domain filtering based on each of the constraints using the constraint filtering algorithm $\textsc{Filter}_C$ for each constraint $C \in \mathcal{C}$ in the CSP encoding. The constraint filtering algorithm for a given constraint identifies variable-value assignments that it can prove will not satisfy the constraint based on the structure of the constraint and the current domains. The extent to which a filtering algorithm removes values is dictated by the level of constraint consistency desired; see Section \ref{sec:consistency}. The output of $\textsc{Filter}_C$ is the filtered domains\footnote{For ease of presentation, we assume the output of $\textsc{Filter}_C$ is all of the updated domains. In practice, this function would only return updated domains for the subset of variables involved in constraint $C$.} and a flag indicating whether the constraint is unsatisfiable ($0$) or potentially satisfiable ($\indeterminate$).

After processing each constraint, the process repeats. As mentioned previously, domain values pruned based on the information in one constraint propagate to another, yielding the ability to perform more domain reduction. There are two cases for the filtering loop (lines $2-14$) to terminate: i) when, after calling the filtering algorithm on all of the constraints, the domains are unchanged, or ii) when a filtering algorithm $\textsc{Filter}_C$ returns ``unsatisfiable'' (a value of $0$). In the latter case, an infeasible constraint is enough to declare the node a dead end and a backtrack is initiated. For the former case, if all the domains are singletons, flag $\flag$ is updated to a value of $1$ indicating a feasible solution has been found. Otherwise, at the current node, the value of $\flag$ is left as indeterminate ($\indeterminate$). The pruned domains are returned and the branching operator, described in the next section, is used to generate a child and continue the search. 

The development of efficient filtering algorithms is of utmost importance to the success of CP. Better logical inference can detect dead ends earlier in the search (thereby pruning away many nodes that would otherwise need to be explored) but usually at the cost of being more expensive (i.e., time consuming) at each node that is explored.
It is therefore useful to classify the extent to which a particular filtering algorithm removes values in order to balance this tradeoff. This topic forms the core discussion in \cref{sec:consistency}.

\subsubsection{Branching Operator}  While some CP searches use naive backtracking (as described above), the paradigm commonly employs alternative branching strategies including (but not limited to) 2-way branching, which creates two children by posting constraints $x_i = d_j$ and $x_i \neq d_j$ \cite{van2006backtracking} for some $d_j \in D_i$, and domain splitting, which creates two children by posting constraints of the form $x_i \leq d_j$ and $x_i > d_j$.\footnote{Domain splitting is a commonly used branching strategy for branch-and-bound approaches for solving integer programs.} Generally, a branch in CP adds a unary constraint (not limited to an equality constraint as in naive backtracking) to the CSP. To preserve the completeness of the search, the total set of branching constraints posted from a node must be exhaustive. 

We can equivalently view branching in CP as an operation which, given a tuple of domains at a node in the search tree, produces a set of child nodes, each associated with a tuple of domains. In this case, instead of a node being defined by the original CSP and the set of branching constraints posted, a node is instead defined by the original variables $X$, constraints $\mathcal{C}$, and the tuple of active domains (where the active domains incorporate all domain reductions from previous branching decisions as well as reductions due to filtering at each node). This formulation will be particularly useful for our discussion of quantum backtracking extensions in \cref{sec:q-backtracking-with-inference}. Formally, we define the branching operator
\begin{equation}
h_c: 2^{\mathcal D} 
\to 2^{\mathcal D}
\end{equation}
as taking the parent's tuple of domains as input and returning the $c$-th child (another tuple of domains) produced by branching. We also define the operator 
\begin{equation}
\numChildren : 2^{\mathcal D} \to \mathbb{Z}^{+}
\end{equation}
as taking the parent's tuple of domains as input and returning the number of children generated by branching from the parent (as determined by the specific branching strategy used). 
Classically, such a function is not typically needed explicitly, because the children are explored sequentially; however, for the quantum algorithms we propose in this work it will be necessary.
To maintain completeness, we require that 
\begin{equation}
\bigtimes_{i=1}^{|X|} D_i
=
\bigcup_{c} 
\bigtimes_{i=1}^{|X|} D_i^{(c)}
\end{equation}
where $D^{(c)}_i$ is the domain of variable $x_i$ in the $c$-th child. 
That is to say, any assignment consistent with the domains of the parent is consistent with the domains of at least one child.

\begin{algorithm}[t]
\SetAlgoLined
\KwData{Domains $\mathcal{D}$}
\KwResult{Filtered domains $\mathcal{D}'$, flag $\flag$}
$\flag \leftarrow \indeterminate$\;
\Repeat{status = False}{
    status $\leftarrow$ False\;  
    \For{$C \in \mathcal{C}$}{
        $\mathcal{D}', \flag \leftarrow \textsc{Filter}_C(\mathcal{D})$\;
        \If {$\flag = 0$}{
            \KwRet{$\mathcal{D}', \flag$}
        }
        \ElseIf{$\mathcal{D} \neq \mathcal{D}'$} {
            status $\leftarrow$ True\;
            $\mathcal{D} \leftarrow \mathcal{D}'$\; 
        }
    }
}
\If{$|D_i| = 1, \forall D_i \in \mathcal{D}'$}{
$\flag \leftarrow 1$\;
}
\KwRet{$\mathcal{D}', \flag$}
 \caption{Propagation function $F$}
 \label{alg:filter-domains}
\end{algorithm}

\subsection{Consistency}\label{sec:consistency}

Concepts of consistency have long played a key role in CP and are fundamentally important to the performance of backtracking search algorithms~\cite{davarnia2019consistency}. In this section, following previous work~\cite{van2006global}, we consider \emph{domain consistency} (also known as \emph{generalized arc consistency}) and \emph{range consistency} but recognize that there are other relevant notions of consistency.~\cite{mackworth1977consistency,davarnia2019consistency}.

\begin{definition}[Domain consistency]
An $m$-ary constraint $C$ with scope $(x_1,\dots,x_m)$ having non-empty domains is called domain consistent iff for each variable $x_i$ and every value ${d}^*_i$ in its domain $D_i$, there is an assignment $(d_1^*, \ldots, d_m^*)$ such that $d^*_j \in D_j$ for all $j \in \{1, \ldots, m\} \setminus \{i\}$ and that satisfies constraint $C$. 
\end{definition}

Intuitively, the constraint is domain consistent if, after assigning any variable to any value in its domain, there exists a set of values to assign to the other variables (from their domains) such that the constraint is satisfied. From this, we can define a domain consistent CSP as follows:

\begin{definition}
[Domain consistent CSP] A CSP is domain consistent iff all of its constraints are domain consistent.
\end{definition}

\begin{example}\label{ex2}
Consider the CSP:\ $X = \left(x_1, x_2, x_3\right)$, with $D_1 = \{1,2\}$, $D_2 = D_3 = \{2,3\}$ and the constraints $x_1 < x_2$ and $x_1 < x_3$. This CSP is domain consistent as each of the constraints is domain consistent.
\end{example}

Establishing domain consistency for some constraints can be, in the worst case, as intractable as solving the global CSP. For this reason there exist weaker forms of consistency, such as range consistency defined as follows \cite{van2006global}.

\begin{definition}[Range consistency] 
An $m$-ary constraint $C$ with scope $(x_1,\dots,x_m)$ having non-empty, real-valued domains is called range consistent iff for each variable $x_i$ and every value ${d}^*_i$ in its domain $D_i$, there is an assignment $(d_1^*, \ldots, d_m^*)$ such that $\min D_j \leq d^*_j \leq \max D_j$ for all $j \in \{1, \ldots, m\} \setminus \{i\}$ and that satisfies constraint $C$. 
\end{definition}

Range consistency is effectively a relaxation of domain consistency, only checking the feasibility of the constraint with respect to the range of domain values instead of the domain itself. As such, filtering for range consistency is never more costly than domain consistency, though the difference in effort depends on the specific constraint involved.
Domain consistency (or another specified level of consistency) for a constraint is achieved via the filtering operator for that constraint. 
As effort spent filtering domains typically results in fewer nodes in the search tree, much of the effort in CP has been in identifying constraints that promote efficient and effective filtering. These special constraints are known as global constraints.

\subsection{Global constraints}

A \emph{global constraint} is a constraint acting on a more-than-constant number of variables that represents a commonly recurring combinatorial substructure~\cite{van2006global}. The motivation for the use of global constraints is twofold. First, the shorthand of the constraint simplifies the high-level modeling task. Second, while an equivalent constraint relationship may be expressed with a combination of simpler constraints, global constraints can strengthen the performance of solvers by maintaining a more global view of the structure of the problem. This often translates to the ability to perform more domain filtering. 

To illustrate this concept, we introduce a concrete example involving the \texttt{alldifferent} global constraint~\cite{van2001alldifferent}. 

\begin{definition}
[\texttt{alldifferent} constraint] The constraint $\texttt{alldifferent}(x_1,\dots,x_k)$ requires that all of the variables in its scope take on different values (i.e., in a solution to the constraint $x_i \neq x_j$ for all $i < j \in \{1,\dots,k\}$).
\end{definition}

The \texttt{alldifferent} constraint captures a commonly recurring substructure, namely that a subset of problem variables need to take on different values. This structure often occurs in timetabling and scheduling problems, for example. In addition to packaging this structure nicely for modeling purposes, the global constraint also increases the amount of inference that can be performed.

\begin{example}
Consider the CSP:\ $X = \left(x_1, x_2, x_3\right)$, with $D_1=D_2=D_3=\{1,2\}$, and the constraints $x_1 \neq x_2$, $x_1 \neq x_3$, and $x_2 \neq x_3$.
Enforcing domain consistency on each constraint does not allow us to perform \emph{any} inference; each constraint is domain consistent and the pruned variable domains remain the same. It follows that the CSP is domain consistent as posed, even though its unsatisfiability is apparent. 
\end{example}

In contrast, consider the impact of using the \texttt{alldifferent} global constraint: 

\begin{example}
Consider the CSP:\ $X = \left(x_1, x_2, x_3\right)$, with $D_1=D_2=D_3=\{1,2\}$, and the constraint $\texttt{alldifferent}(x_1, x_2, x_3)$.
Enforcing domain consistency on the \texttt{alldifferent} constraint would quickly return infeasible, as assigning $x_1 =1$ does not permit a set of values for $x_2$ and $x_3$ that would satisfy the constraint, nor does $x_1 = 2$, thus emptying the domain store of $x_1$. As such, the constraint (and thus the CSP) is inconsistent.
\end{example}

Evidently, it is beneficial to represent a relationship of difference over a set of variables as an \texttt{alldifferent} constraint. 
In general, global constraints are proposed for substructures that permit enhanced filtering, not just to facilitate a more concise expression of CSPs. The library of available global constraints is extensive\footnote{A comprehensive global constraint catalogue is available at: https://sofdem.github.io/gccat/} and useful for modeling a wide array of problems \cite{beldiceanu2007global}. The success of these global constraints, however, is largely tied to the efficiency of their underlying filtering algorithm.

The worst-case complexity of filtering algorithms for domain consistency can be polynomial or exponential (i.e., as intractable as the problem being solved), depending on the constraint. To mitigate this, weakened forms of consistency can be accepted, or the filtering algorithm can simply be terminated prior to achieving domain consistency in favor of branching in the search tree. Much of the research effort in the CP community revolves around the design of algorithms that achieve a given level of consistency with improved worst-case complexity over existing methods (e.g., for scheduling problems with resource constraints~\cite{mercier2008edge,vilim2009edge}).  

\subsection{CP modeling and solving: Sudoku} \label{sec:sudoku}

\begin{figure}
\centering
        \includegraphics[width=0.25\textwidth]{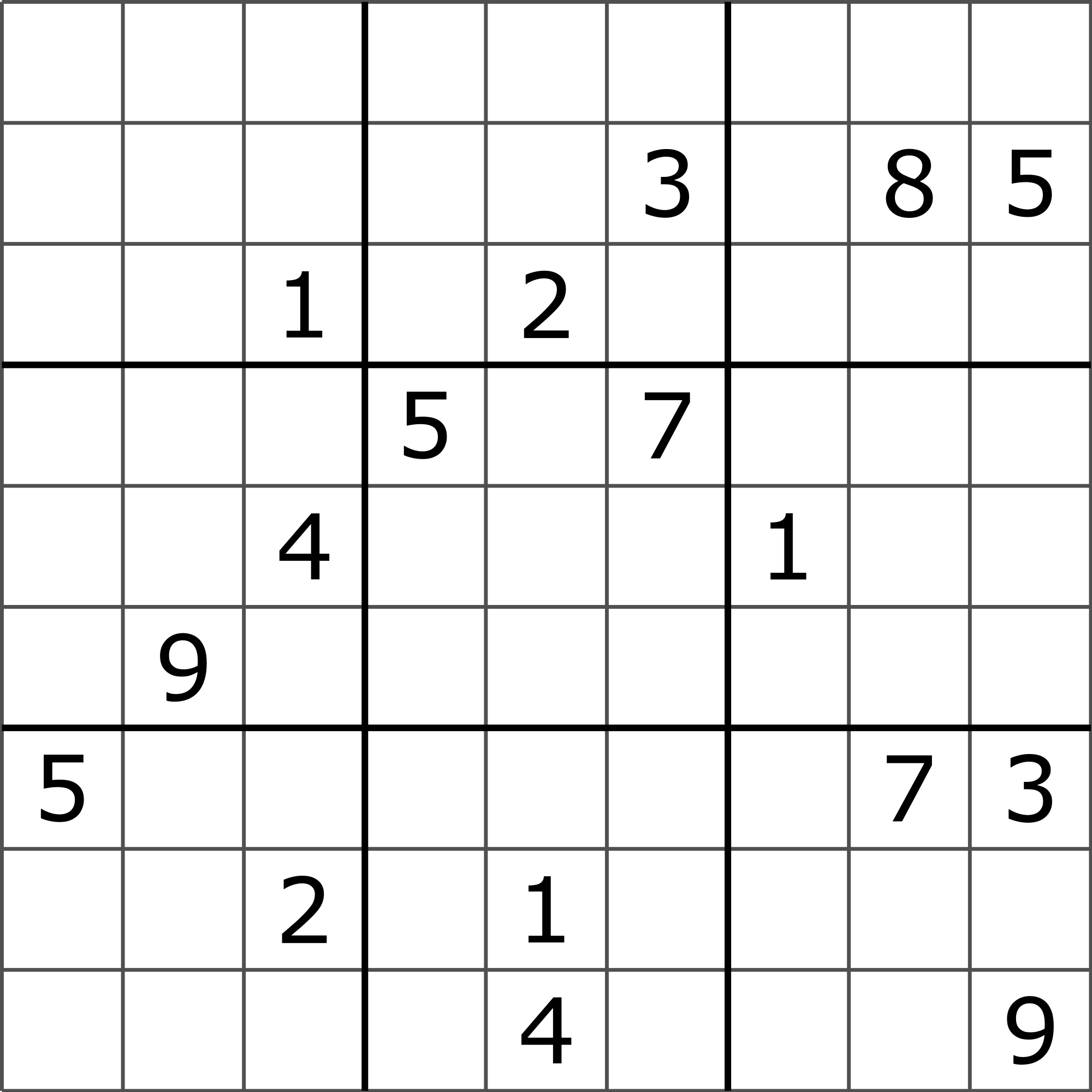}
    \caption{Sudoku problem instance with $9 \times 9$ cells.}
    \label{fig:sudoku}
\end{figure}

To illustrate CP's branch-and-infer search, consider a CP model for the popular puzzle game Sudoku~\cite{simonis2005sudoku}. 
An example Sudoku puzzle is visualized in Figure~\ref{fig:sudoku}. 
A solution to the puzzle assigns a number ranging 1--9 to each cell such that each row, column, and $3 \times 3$ sub-grid contains all of the digits from 1 through 9. 
The decision problem related to solving general Sudoku puzzles on $n^2 \times n^2$ grids is NP-complete \cite{yato2003complexity}. In this example, $n=3$.

One option for modeling the problem with CP uses integer decision variables of the form $x_{i,j} \in \{1,\dots,9\}$, representing the value placed in the cell of row $i$, column $j$. Since it is a satisfiability problem, we do not have an objective function. Let $I = \{1,\dots,9\}$. The CP model includes the 
variables
$
x_{i,j} \in I, \forall i,j \in I
$
subject to the following constraints:
\begin{align}
& x_{3,3} = 1, x_{3,5} = 2, x_{2,6} = 3, \ \text{etc.} & \label{cp:clues} \\
& \texttt{alldifferent}(x_{i,1},\dots,x_{i,9}),  \forall i \in I & \label{cp:rows} \\
& \texttt{alldifferent}(x_{1,j},\dots,x_{9,j}), \forall j \in I  & \label{cp:cols}\\
& \texttt{alldifferent}(x_{i,j}, x_{i,j+1}, x_{i,j+2},  \nonumber\\
& \quad \quad \quad \quad \quad \quad \quad x_{i+1,j},x_{i+1,j+1}, x_{i+1,j+2}, \nonumber \\
& \quad \quad \quad \quad \quad \quad \quad x_{i+2,j}, x_{i+2,j+1}, x_{i+2,j+2}), \forall i,j \in \{1,4,7\} \label{cp:cells}
\end{align}
Constraint~\eqref{cp:clues} embeds the ``clues'' of the problem, fixing them to take on the values given in the instance. Constraint~\eqref{cp:rows} ensures that each cell in a given row takes on a different value, and Constraint~\eqref{cp:cols} does the same for columns. Constraint~\eqref{cp:cells} enforces that each $3 \times 3$ sub-grid must have all different values. We note that, however, this is not the only possible model. For example, we could have used binary decision variables $x_{i,j,k}$, taking on a value of 1 if value $k \in \{1,\dots,9\}$ is assigned to the cell in row $i$, columnn $j$, and 0 otherwise. Of course, the set of constraints using this variable definition would be different as well. \cref{sec:other-models} provides modeling examples for other problems using CP.

With the model in hand, we can use CP's branch-and-infer search to solve the Sudoku instance. The search would start by invoking the propagation function to perform root node filtering over the model constraints. For example, enforcing domain consistency on $\texttt{alldifferent}(x_{1,1}, x_{1,2}, \dots, x_{1,9})$ would not change the variable domains (as row 1 is blank). However, enforcing domain consistency on $\texttt{alldifferent}(x_{2,1}, x_{2,2}, \dots, x_{2,9})$ would prune a number of values (e.g., the value 3 would be pruned from the domain of $x_{2,1}$) as the hints in this row permit some inference. After the model achieves the desired level of consistency, the search branches by, for example, assigning a value to a variable (i.e., guessing a value assignment to a cell), before repeating the process. 

The CP model described above has $9 \times 9 = 81$ decision variables, each with an initial domain of size $|I| = 9$, in the worst case. Each of the \texttt{alldifferent} filtering subproblems, however, only involves nine variables. While representing the entire problem on a quantum chip may be prohibitive, the filtering subproblems for the \texttt{alldifferent} constraints can more readily take advantage of early fault-tolerant quantum hardware. 

Within CP's branch-and-infer tree search, an improvement in the efficiency of global constraint filtering allows for domain values to be pruned faster. 
While this does not change the number of nodes explored before finding a solution, which is usually the dominant factor in the runtime, it does reduce the per-node time and therefore has a significant positive impact on the efficiency of solving problems in practice.
This phenomenon has been demonstrated in similar paradigms such as integer programming (IP) where improvements in solving linear programming (LP) relaxations has had an orders-of-magnitude impact on the performance of IP solvers over the past few decades~\cite{bixby2012brief}.

%% file: sections/3-related-work.tex
\section{Related work}\label{sec:related-work}

In this section we identify work immediately relevant to this paper. In Sections~\ref{sec:qram}, \ref{sec:filtering}, and \ref{sec:tree-search} we present additional background in the context of our results. Similarly titled work by Di Pierro et al. introduced a formal language for expressing quantum computations that they called ``quantum constraint programming''~\cite{dipierro2001quantum}; that is, they apply the \emph{modeling} aspect of CP to quantum computations, whereas here we employ a quantum version of the classical CP approach to \emph{solving} problems. 

This work investigates the quantum acceleration of CP, at the levels of both inference and search. Our explorations make use of two important quantum algorithms: quantum search and phase estimation. For the former, Grover's algorithm famously shows that a target element can be found in an unstructured list of $N$ elements with only $O(\sqrt{N})$ quantum queries, which gives a realizable quadratic speedup over classical black-box search (requiring $N$ queries in the worst case) when the oracle for accessing the list can be queried or implemented efficiently~\cite{grover1996fast}. Phase estimation is a quantum algorithm used to estimate the eigenvalues of a unitary~\cite{cleve1998quantum}. 

Quantum algorithms for solving various graph problems have seen considerable progress in recent years. These algorithms are often studied in the quantum query model, where the algorithm accesses the graph through a quantum query operation as we detail in Section~\ref{sec:qram}. Previous work provides lower and upper bounds for the bounded-error quantum query complexity of various graph problems, including connectivity, minimum spanning tree, and single-source shortest path~\cite{berzina2004quantum,durr2006quantum,furrow2006panoply,cai2007quantum,lin2014upper}. Particularly relevant work has investigated the quantum query complexity of 
matching problems~\cite{dorn2009quantum,lin2014upper,zhang2005power,ambainis2006quantum,beigi2020quantum,kimmel2020query}, showing speedups over classical query-model algorithms, and in some cases~\cite{ambainis2006quantum,dorn2009quantum} explicitly improved time complexities, as discussed further in Section \ref{sec:max-matching}.

Similarly, there have been efforts to develop quantum algorithms for constraint satisfaction and search. Quantum search has been extended to problems within mathematical programming, such as semidefinite programming~\cite{brandao2017quantum,van2020quantum} and the acceleration of the simplex method~\cite{nannicini2019fast}. The latter, in a similar fashion to this work, uses quantum search to accelerate the subroutines of the simplex method (e.g., variable pricing). There also exist recent efforts to use algorithms based on quantum search and phase estimation to speed up tree search methods including branch-and-bound~\cite{montanaro2020quantum} and backtracking search~\cite{montanaro2015quantum,jarret2018improved,ambainis2017quantum}; we discuss the latter in more detail in~\cref{sec:tree-search}. 
See in particular \cite[Sec.~1.3]{montanaro2015quantum} for a historical overview. 

%% file: sections/4-quantum-access.tex
\section{Quantum resources and data access}\label{sec:qram}

In this paper, we propose quantum algorithms and quantum subroutines to solve problems whose instance specification and solution are classical information.
Many of these quantum algorithms require quantum access to classical data that enables computation on a superposition of classical data.
This section synthesizes the relevant ideas from the literature as background for later sections.
For example, suppose that an instance is specified by a function $f: W \to Y$, where $Y = {\{0,1 \}}^r$ and elements of $Y$ can be viewed as integers with addition modulo $|Y| = 2^r$.
By ``quantum access'', we mean the ability to call the unitary
\begin{equation}\label{eqn:quantum-oracle}
U_{f} \ket{w} \ket{y} = \ket{w} \ket{y + f(w) \bmod{|Y|}}
\end{equation}
on an arbitrary state $\sum_{w \in W, y \in Y} \tau_{w, y} \ket{w} \ket{y}$.
We will refer to the first register $\ket{w}$ as the query register and the second register $\ket{y}$ as the database register.

Quantum algorithms are assessed with respect to several different metrics: query complexity, time complexity in the oracle model, and time complexity in the gate model.
In the quantum oracle model, we are given access to a unitary such as that expressed in~\cref{eqn:quantum-oracle} (the ``oracle'').
Query complexity counts the number of queries (calls to the oracle) the algorithm uses.
Time complexity in the oracle model additionally counts the number of primitive (two-qubit) gates, counting each query as a single gate.
In the gate model, there is no oracle; everything must be expressed in terms of primitive gates, including the query unitary.
(While the oracle model only exists in theory, it can be extremely useful, especially for proving lower bounds, and in some cases directly leads to improved  gate-model time complexities.)
The time complexity is usually the number of primitive gates (i.e., the size of the circuit)\footnote{When using circuit complexity measures, whether in the classical or quantum setting, care must be taken that uniformity conditions are met so that complexity is not hidden in the circuit specification~\cite{arora2009computational}.}, but can also be the depth of the circuit if parallelization is permitted.

The quantum time complexities stated in this work are in terms of the depth of quantum circuits consisting of two-qubit gates, but with parallelization only within the parts of the circuits that implement the queries.
The classical run-times reported, however, are in a higher-level model that neglects non-dominant logarithmic factors associated with bit-level operations; a lower-level model, such as classical circuits consisting of binary gates, would include such additional factors. To abstract away such implementation details, we consider only the dominant factors when comparing classical and quantum.

There are two main ways of implementing the query in the gate model. 
The first can be used when the data specified by $f$ can be efficiently and uniformly computed classically,  
in which case one can use an explicit, efficient
quantum circuit computing the function $f$.
For example, in applying Grover's algorithm to a problem in NP, the function $f$ is simply the verifier, which by definition can be efficiently computed classically, and thus also by an efficient quantum circuit. For any classical circuit with $t$ gates on $n$ bits, one can construct a reversible version (and thus suitable for quantum circuit implementation) with just $O(t^{1+o(1)})$ gates on $O(n \log t)$ bits. See~\cite[Ch. 6]{rieffel2011quantum} for details.
For unstructured data, the second method, which we employ in this work, is called quantum random access memory (QRAM)~\cite{qram}.

QRAM is a data structure that allows quantum queries of the form expressed by Eq.~\eqref{eqn:quantum-oracle}.
Broadly speaking there are two classes of QRAM:\@ i) a speculative, yet plausible, special-purpose physical hardware analogous to classical RAM~\cite{jiang2019experimental}, and ii) circuit-based QRAM~\cite{dimatteo2020fault,arunachalam2015robustness}.
In the former, the number of qubits required is linear in the database size, $O(|W|\log |Y|)$, and query calls occur in logarithmic time  $O(\log |W| + \log |Y|)$. The latter circuit-based QRAM, on the other hand, is more flexible in terms of quantum resources supporting trade-offs between circuit width (number of qubits) and depth (number of quantum gates).
Circuit QRAM can be implemented \emph{implicitly} using $O(\log |W| + \log |Y|)$ total qubits with circuit depth $O(|W|\log |Y|)$, or \emph{explicitly} using $O(|W|\log |Y|)$ total qubits and $O(\log |W| + \log \log |Y|)$ depth.\footnote{
Given $\log |Y|$ copies of the index $w \in W$, each bit of the output $y \in Y$ can be queried in parallel in $O(\log |W|)$ depth.
The $\log |Y|$ copies can be made (and unmade) in $O(\log \log |Y|)$ depth.
}
Because gates can act in parallel, the total number of primitive gates can still be linear in the database size, even with logarithmic depth.
\footnote{
While practical quantum computers are expected to employ significant parallelization, the circuit model we use has no constraints on the geometry of the gates. That is, two gates can act in parallel so long as they act on disjoint sets of qubits.
Such non-local parallelization, while plausible (more so because we only require it for a specific type of memory circuit) may be difficult to realize in practice.
}
Henceforth, by ``QRAM'' we will mean either the special-purpose hardware or explicit circuit-based variants,
with linear space, logarithmic access time (circuit depth, in the latter case), and linear initialization time (but logarithmic update time). In both variants, the contents of the database are stored explicitly in memory. In some cases we will require this memory content to be in a superposition, as further detailed in Section~\ref{sec:q-backtracking-with-inference}.

We are primarily interested in quantum query access to a directed or undirected graph $G = (V, E)$, with $n=|V|$ vertices and $m=|E|$ edges.
We consider two models for accessing the graph: the adjacency matrix model (i.e., the ``matrix'' model) and the adjacency list model (i.e., the ``list'' model).

\paragraph{Adjacency matrix model.} Let $A \in {\{0, 1\}}^{|V| \times |V|}$ represent the adjacency matrix of $G$ such that $A_{u, v} = 1$ iff $(u, v) \in E$.  In the matrix model, the query is defined by
\begin{equation}\label{eqn:adjacency-query}
U_A \ket{u, v} \ket{b} = \ket{u, v} \ket{b \oplus A_{u, v}}
\end{equation}
for all $u, v \in V$ and $b \in \{0, 1\}$. 
A QRAM implementing \cref{eqn:adjacency-query} can be initialized in time $O({|V|}^2)$ and 
queried in time $O(\log|V|)$. 

\paragraph{Adjacency list model.} Let $\degree_v$ be the degree of vertex $v \in V$ and $N_v : [\degree_v] \rightarrow [n]$ be an array with the neighbors of vertex $v$. Then, in the list model, we can query
\begin{equation}\label{eqn:list-query}
U_{\ngbrs_v} \ket{i} \ket{j} = \ket{i} \ket{j + \ngbrs_v(i) \bmod{\degree_v}}
\end{equation}
where  $i \in [\degree_v]$, and $j \in [\degree_v]$, where $\ngbrs_v(i)$ is the $i$th neighbor of vertex $v$. For each vertex $v$, a QRAM implementing \cref{eqn:list-query} can be initialized in time $O(\degree_v \log n)$ and called in time $O(\log n)$. 
The overall initialization time for the graph is thus at most $O(m \log n)$, though we will often need to initialize only the relevant portions of the model at a given point in an algorithm, and in many cases this too can be parallelized further.

In this paper, we primarily use the list model due to its superior performance in the application we consider.

%% file: sections/5-filtering.tex
\section{Quantum-accelerated global constraint filtering}\label{sec:filtering}

In this section we detail a quantum-accelerated filtering algorithm for domain consistency of the \texttt{alldifferent} constraint. 
In~\cref{sec:classical-filtering}, we review \Regin's classical filtering algorithm for the \texttt{alldifferent} constraint.
In~\cref{sec:max-matching}, we explain how \Dorn's quantum algorithm for maximum matching can be used to speed up the costliest part of \Regin's algorithm.
In~\cref{sec:remove-edges}, we explain how several quantum algorithms can be combined to speed up the less costly parts of \Regin's algorithm.
In~\cref{sec:generalized}, we explain how Cymer's general filtering framework allows for using quantum maximum matching to filter other global constraints whose domain-consistency algorithms are structurally similar to that for \texttt{alldifferent}.

\input{sections/5-filtering/all_different.tex}
\input{sections/5-filtering/generalized.tex}

%% file: sections/5-filtering/all_different.tex
\subsection{The \texttt{alldifferent} constraint}

The proposed quantum subroutines accelerate the classical algorithm of R\'{e}gin~\cite{regin1994filtering} for filtering the \texttt{alldifferent} constraint. We note that more recent work has investigated techniques for optimizing R\'{e}gin's algorithm in practice; however, these do not improve upon its worst-case time complexity~\cite{gent2008generalised,zhang2018fast}.

\subsubsection{Classical filtering algorithm}\label{sec:classical-filtering}
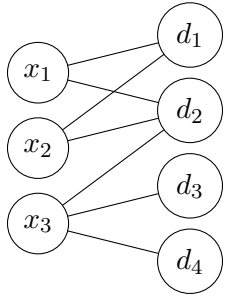
\begin{figure}
\centering
        \centering
     \begin{tikzpicture}
      [every node/.style = {draw, circle, fill=white}]
    \draw (0,2)--(2,2.5);
    \draw (0,2)--(2,1.5);
    \draw (0,1)--(2,2.5);
    \draw (0,1)--(2,1.5);
    \draw (0,0)--(2,1.5);
    \draw (0,0)--(2,0.5);
    \draw (0,0)--(2,-0.5);
    \node at (0,0) {$x_3$};
    \node at (0,1) {$x_2$};
    \node at (0,2) {$x_1$};
    \node at (2,2.5) {$d_1$};
    \node at (2,1.5) {$d_2$};
    \node at (2,0.5) {$d_3$};
    \node at (2,-0.5) {$d_4$};
    \end{tikzpicture}
    \caption{Example bipartite variable-value graph $G$. }
    \label{fig:bipartite-value}
\end{figure}

The classical filtering algorithm for \texttt{alldifferent} begins by constructing a bipartite variable-value graph~\cite{regin1994filtering}, as illustrated in Figure~\ref{fig:bipartite-value}.
The example visualized involves variables and domains $x_1 \in D_1=\{d_1,d_2\}$, $x_2 \in D_2=\{d_1, d_2\}$, $x_3 \in D_3=\{d_2, d_3, d_4\}$. 
One solution to the constraint $\texttt{alldifferent}(x_1,x_2,x_3)$ would be $x_1 = d_1,\, x_2=d_2,\, x_3=d_3$, but there are other possibilities as well. A domain-consistency filtering algorithm for this constraint seeks to remove values in each domain that cannot participate in a solution to the constraint. For this example, $x_3=d_2$ is an assignment that will never be feasible and thus $d_2$ should be pruned from the domain of $x_3$. 

Recall the notation from Section \ref{sec:csp}. We define the bipartite variable-value graph as $G=(X, V, E)$, with vertices $X\cup V$ and edges $E$. Each edge in the graph represents a variable-value pair. In this case, $V = \bigcup_i D_i$ is the set of unique domain values. Such a graph has $n=|X|+|V|$ vertices and $m=|E|=\sum_i |D_i| \leq |X| |V|$ edges, with $|E|\geq |V|$.

\begin{algorithm}[ht]
\SetAlgoLined
\KwData{Variables $X$ and domains $\mathcal{D}$}
\KwResult{False if no solution, otherwise filtered domains }
Build $G=(X, V, E)$\;
$M \leftarrow \textsc{FindMaximumMatching}(G)$\;
\If{$|M| < |X|$}{
  \KwRet{False}
  }
$\mathcal{D}' \leftarrow \mathcal{D} \setminus \textsc{RemoveEdges}(G, M)$\;
\KwRet{$\mathcal{D}'$}
 \caption{\texttt{alldifferent} filtering~\cite{regin1994filtering}}
 \label{alg:alldifferent}
\end{algorithm}

With $G$, the filtering of \texttt{alldifferent} proceeds as detailed in Algorithm~\ref{alg:alldifferent}. The algorithm consists of two primary subroutines: \textsc{FindMaximumMatching}, which finds a matching of maximum size (i.e., one with the most edges) in $G$, and \textsc{RemoveEdges}, which identifies edges in $G$ that can never participate in a maximum matching. If \textsc{FindMaximumMatching} returns a matching $M$ whose number of edges $|M|<|X|$, the constraint cannot be satisfied and the algorithm terminates.
If a maximum matching exists with $|M| = |X|$, the algorithm prunes domains based on the output of \textsc{RemoveEdges}. The result is a set of pruned variable domains such that the constraint is domain consistent. 

The \textsc{FindMaximumMatching} subroutine bears the brunt of the computational complexity~\cite{van2001alldifferent}.
For our purposes, we only invoke this subroutine when $|V| \geq |X|$ (as the case $|X| > |V|$ is clearly infeasible). Long-standing previously state-of-the-art classical algorithms for finding maximum matchings run in $O(m\sqrt{n})$ time; the algorithm of Hopcroft and Karp (HK) is for bipartite graphs~\cite{hopcroft1973n}, while the algorithm of Micali and Vazirani (MV) applies to general graphs~\cite{micali1980v,vazirani2012simplification}. Given any initial matching, these algorithms operate in phases, where each phase of the algorithm looks to find a matching of greater size. The runtime of each phase is $O(m)$, and $O(\sqrt{|M|}) = O(\sqrt{n})$ phases are required~\cite{hopcroft1973n}, where $|M|$ is the size of the maximum matching. In terms of the variable-value graph properties, since $|M|$ is bounded by $|X|$, these algorithms take $O(|E|\sqrt{|X|})$ time.

Following the algorithms of HK and MV, a randomized $O(n^{\omega})$-time algorithm~\cite{mucha2004maximum,ibarra1981deterministic} was proposed for bipartite graphs, where $\omega$ corresponds to the classical asymptotic cost of matrix multiplication; the best upper bound known on $\omega$ is approximately $2.373$
~\cite{alman2021refined}. In terms of the variable-value graph properties, this algorithm takes $O(|X|^{\omega-1}|V|)$ time~\cite{ibarra1981deterministic}. Alt et al.\ then proposed an $O(n^{3/2} \sqrt{m/\text{log} \ n})$ algorithm~\cite{alt1991computing}. Each of these algorithms offer modest improvements over HK for dense graphs.

Finally, very recent work using interior-point methods and dynamic graph algorithms have led to an 
$\tilde{O}(m + n^{3/2})$-time classical algorithm~\cite{brand2020bipartite} for finding maximum matchings in bipartite graphs, offering the first significant improvement over HK. The algorithm leverages fast linear system solvers to provide near-linear asymptotic complexity for moderately dense graphs. In terms of the variable-value graph properties, with $n = O(|V|)$, their algorithm runs in $\tilde{O}(|E|+|V|^{3/2})$ time.

In order to remove edges which participate in no maximum matching, and thus cannot satisfy the constraint, \textsc{RemoveEdges} finds strongly connected components (SCCs) in a directed transformation of $G$ using Tarjan's $O(n +m)$ algorithm~\cite{tarjan}.
While this subroutine is not the dominant contribution to the computational cost of \texttt{alldifferent} filtering, its acceleration can still be valuable in practice.

In the remainder of this section we provide details for the \textsc{FindMaximumMatching} and \textsc{RemoveEdges} subroutines that accelerate the filtering of the \texttt{alldifferent} constraint. For the former, we summarize existing quantum algorithms for finding maximum matchings in graphs, noting the complexity of the state of the art. For the latter, we combine a number of quantum graph algorithms, including an adaptation of work that identified strong connectivity in graphs~\cite{durr2006quantum}, into a quantum subroutine that improves over the classical algorithm in some cases.  

\subsubsection{Subroutine: Finding a maximum matching}\label{sec:max-matching}

The essence of a quantum-accelerated filtering algorithm 
for \texttt{alldifferent}
is simple: use a quantum algorithm to solve the maximum matching problem.
Recent work proposed a series of algorithms for finding maximum matchings in the quantum query model~\cite{ambainis2006quantum,dorn2009quantum}. To the authors' knowledge, and excluding an earlier version of this work \cite{booth2020quantum}, 
these results have never been linked to accelerating global constraint filtering in CP.

In the list model (see Section \ref{sec:qram}), an initially proposed quantum algorithm is capable of finding maximum matchings in $O(n\sqrt{m+n} \log^2 n)$ time~\cite{ambainis2006quantum}, while a second, improved algorithm runs in $O(n\sqrt{m} \log^2 n)$ time~\cite{dorn2009quantum}.\footnote{It is known that the quantum query complexity must be $\Omega(n^{3/2})$, even for bipartite graphs~\cite{berzina2004quantum,zhang2005power}.} 
The latter algorithm, proposed by \Dorn, improves over both existing deterministic and randomized algorithms for the majority of parameter values, and follows the classical MV algorithm for finding maximum matchings in general graphs~\cite{micali1980v}, but accelerates its primary subroutines with quantum search. 

In \Dorn's time-complexity result, the first $\log$ factor is due to repetitions of quantum search required to produce an overall constant success probability bound\footnote{By ``success probability'' we mean the probability of the algorithm returning a correct answer.} for the algorithm~\cite{dorn2009quantum}. Each individual instance of quantum search may provide an incorrect answer with a constant probability~\cite{grover1996fast}, and the overall algorithm uses $\poly(n)$ quantum searches. To ensure the probability of the overall algorithm producing an incorrect answer is less than $O(1/\poly(n))$, each instance of quantum search is repeated $O(\log n)$ times for a success probability of at least $1-O(1/\poly(n))$.
The implications of these probabilties are discussed more in the context of backtracking search in \cref{sec:classical-tree-search}. The second $\log$ factor is due to the cost of the (QRAM) queries, as discussed previously in \cref{sec:qram}, and other low-level implementation costs.

As in HK, \Dorn's algorithm for bipartite graphs consists of $O(\sqrt{|X|})$ phases, each of which takes $O(\sqrt{nm}) = O(\sqrt{|V||E|})$ time. Thus, in terms of the variable-value graph properties, the time complexity of \Dorn's algorithm is $O(\sqrt{|X||V||E|}\log^2 |V|)$, for a constant $\Omega(1)$ success probability~\cite{dorn2009quantum}.
This indicates an improvement by a factor on the order of $\sqrt{|E|/|V|}$ over HK and MV, up to polylogarithmic terms. A similar analysis yields an improvement by a factor on the order of $\sqrt{|X|^{2\omega-3}|V|/|E|}$ over the randomized matrix multiplication algorithm discussed above.

\Dorn's algorithm also offers an improvement over Brand et al.'s interior-point-based method (IPM)~\cite{brand2020bipartite} for the regime where $|E| = O(|V|^{3/2})$, yielding an improvement factor on the order of $|V|/\sqrt{|X||E|}$ and up to polylogarithmic terms. Within this regime, when $|X|= O(\sqrt{|V|})$ there is always an improvement (i.e., the improvement factor is $\geq 1$) while $|X| = \Omega(\sqrt{|V|})$ yields an improvement only if $|X||E|=O(|V|^2)$.

A quantum algorithm for maximum bipartite matching with query complexity O($n^{3/4}\sqrt{m}$) in the list model is shown in~\cite{beigi2020quantum}, with the same complexity recently obtained for general graphs in~\cite{kimmel2020query}. In each case, it remains an open problem to obtain similarly improved time complexities; see, e.g., \cite[Sec. 4]{kimmel2020query}.

In addition to accelerating the filtering of \texttt{alldifferent}, \Dorn's algorithm (and any improved quantum algorithms of the future) for finding maximum matchings~\cite{dorn2009quantum} can play a crucial role in the acceleration of domain-consistency algorithms for a broader family of global constraints, as discussed in Section \ref{sec:generalized}.

\subsubsection{Subroutine: Removing edges}\label{sec:remove-edges}

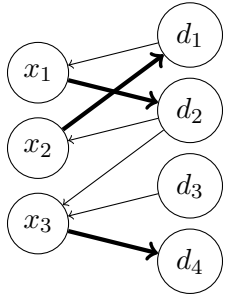
\begin{figure}
\centering
     \begin{tikzpicture}
      [every node/.style = {draw, circle, fill=white}]
      
    \node (x3) at (0,0) {$x_3$};
    \node (x2) at (0,1) {$x_2$};
    \node (x1) at (0,2) {$x_1$};
    \node (v1) at (2,2.5) {$d_1$};
    \node (v2) at (2,1.5) {$d_2$};
    \node (v3) at (2,0.5) {$d_3$};
    \node (v4) at (2,-0.5) {$d_4$}; 
      
    \draw [->, ultra thick, to path={--(\tikztotarget)}] (x1) edge (v2);
    \draw [->, to path={--(\tikztotarget)}] (v1) edge (x1);
    \draw [->, to path={--(\tikztotarget)}] (v2) edge (x2);
    \draw [->, ultra thick, to path={--(\tikztotarget)}] (x2) edge (v1);
    \draw [->, to path={--(\tikztotarget)}] (v2) edge (x3);
    \draw [->, to path={--(\tikztotarget)}] (v3) edge (x3);
    \draw [->, ultra thick, to path={--(\tikztotarget)}] (x3) edge (v4);
    \end{tikzpicture}
    \caption{
        Directed graph $G_M$, a transformation of the variable-value graph $G$. 
        Bold arcs indicate matching $M$, with cardinality $|X|$. 
        Right-facing edges are those in $M$. Left-facing edges are those in $G \setminus M$. In this graph, $d_3$ is an unmatched vertex.}
    \label{fig:bipartite-directed}
\end{figure}

\begin{algorithm}[t]
\SetAlgoLined
\KwData{Bipartite graph $G = (X,V,E)$ and matching $M$}
\KwResult{Set of edges to prune}
$G_M \leftarrow \textsc{DirectGraph}(G,M)$\;
$E_{\mathrm{used}} \leftarrow \textsc{FindSimplePaths}(G_M)$\;
$\mathcal{S} \leftarrow \textsc{FindSCC}(G_M)$\;
\KwRet{$\textsc{IdentifyEdges}(G, M, E_{\mathrm{used}}, \mathcal{S})$}\;
 \caption{$\textsc{RemoveEdges}(G,\,M)$}
 \label{alg:remove-edges}
\end{algorithm}

If a maximum matching is found such that $|M|=|X|$, Algorithm \ref{alg:alldifferent} proceeds to initiate the \textsc{RemoveEdges} subroutine. The subroutine leverages properties formulated by Berge \cite{berge1973graphs} that describe necessary and sufficient conditions for an edge to be involved in some maximum matching. If an edge does not participate in any maximum matching, the edge can be pruned. Instead of applying Berge's conditions directly, the problem has been previously translated into a search for edges in directed simple paths and strongly connected components (SCCs) in a directed transformation of the graph~\cite{regin1994filtering,van2001alldifferent}. We describe the main steps of the \textsc{RemoveEdges} subroutine, detailed by Algorithm \ref{alg:remove-edges}, as follows.

The input to the \textsc{RemoveEdges} subroutine is the variable-value graph $G$ and a matching $M$. In \textsc{DirectGraph}, the edges in $G$ are directed depending upon whether or not they are in the matching $M$, producing directed graph $G_M$. Edges in the matching are directed from variables to values (``right-facing'') and the remaining edges from values to variables (``left-facing''), as shown in Figure \ref{fig:bipartite-directed}. 

Using this graph, \textsc{FindSimplePaths} is used to identify all edges reachable from \emph{unmatched} vertices, i.e., those not involved in the matching $M$. This is achieved by a simultaneous breadth-first search (BFS) from the unmatched vertices, where any edge traversed is marked as ``used''. The output of \textsc{FindSimplePaths} is a set $E_{\mathrm{used}}$ of used edges, and the time complexity is $\Theta(|E_{\mathrm{used}}|)$; 
in the worst case, $|E_{\mathrm{used}}| = \Omega(m)$, but it is typically much smaller. 
In the example of Figure \ref{fig:bipartite-directed}, edges $(d_3,x_3)$ and $(x_3,d_4)$ will be explored by the BFS and thus marked ``used''. These are then added to set $E_{\mathrm{used}}$.

\begin{definition}[Strongly connected component (SCC)] A strongly connected component of a directed graph $G$ is a maximal set of vertices $S$ such that for every pair of vertices $u, v \in S$, there is a directed path from $u$ to $v$ and from $v$ to $u$.
\end{definition}

Next, \textsc{FindSCC} is used to find SCCs in $G_M$. Any edge in an SCC is in some maximum matching and cannot be pruned~\cite{regin1994filtering}.  Tarjan's algorithm can be used to find SCCs with time complexity $O(n+m)$~\cite{van2001alldifferent,tarjan}. In the example of Figure~\ref{fig:bipartite-directed}, vertices $\{x_1,x_2,d_1,d_2\}$ form an SCC while the remaining vertices form trivial SCCs on their own. The output of \textsc{FindSCC} provides a mapping $\mathcal{S} : X\cup V \rightarrow \{1,\dots, N_s\}$ from vertices to the SCC they belong to, where $N_s$ is the number of SCCs in $G_M$.

Finally, \textsc{IdentifyEdges} is used to remove any edges i) not in $M$, ii) not in $E_{\mathrm{used}}$, and iii) not connecting vertices in an SCC. This step has time complexity $O(m)$ as it iterates over all edges in $G_M$. In our running example of Figure~\ref{fig:bipartite-directed}, only edge $(x_3,d_2)$ will be removed, corresponding to the value $d_2$ being pruned from the domain of variable $x_3$.

To summarize, the classical implementation of \textsc{RemoveEdges} has time complexity $O(n+m)$ from the lower-level subroutines \textsc{FindSimplePaths}, \textsc{FindSCC}, and \textsc{IdentifyEdges}. In terms of the variable-value graph properties, this is $O(|E|)$, since $|E|\geq |V|$. Though \textsc{FindSimplePaths} is already asymptotically optimal in the classical case, as the run-time is linear in the size $|E_{\mathrm{used}}|$ of its output, the remaining subroutines can be accelerated using quantum search. 
Specifically, our proposed quantum algorithm improves the time complexity of \textsc{RemoveEdges} to $\tilde{O}\left( |E_{\mathrm{used}}| + \sqrt{|V||E|} + \sqrt{|E||R|} \right)$ (where $R$ is the set of edges that are removed). 
In the worst case, and up to logarithmic factors, this matches the classical complexity of $O(|E|)$, but in the best case, as discussed below, a $\sqrt{|E|/|V|}$ improvement factor can be obtained. In the remainder of this section, we describe how to achieve this. Our approach uses a quantum analog of Tarjan's algorithm, \textsc{Q-FindSCC}, as detailed in Appendix \ref{sec:q-tarjan}.

\paragraph{Initialization.}
Let $M(v)$ be the vertex adjacent to vertex $v$ in matching $M$, where $M(v) = \varnothing$ indicates $v$ is not in the matching. Let $\delta_{M,v}$ be the out-degree of $v$ and $N_{M, v}(i)$ the $i$th out-neighbor of $v$ in $G_M$. To use \textsc{Q-FindSCC}, we need to provide quantum access to $G_M$.
Specifically, we need to provide access to 
\begin{equation}
U_{N_{M, v}} \ket{i} \ket{0} \mapsto \ket{i} \ket{N_{M, v}(i)},
\end{equation}
for all vertices $v \in X \cup V$.
For variable vertices $v \in X$, $\delta_{M,v} = 1$, and so constructing $U_{N_{M, v}}$ is trivial. However, whenever it is called as part of a search, we can simply return the appropriate value classically\footnote{
We assume here that every variable vertex $v \in X$ is matched in $M$, which will always be the case when \textsc{RemoveEdges} is used in \texttt{alldifferent} filtering.
When that's not the case, $U_{N_{M, v}}$ is even simpler: $\delta_{M, v} = 0$.
}.
For unmatched value vertices $v \in V$, $U_{N_{M, v}} = U_{N_{v}}$; i.e., it is the same as that for the undirected graph $G$.
For matched value vertices $v \in V$, $\delta_{M,v} = \delta_{v} - 1$, where $\delta_v$ is the degree of $v$ in the undirected graph $G$.
$U_{N_{M, v}}$ can be implemented by calling $U_{N_{v}}$ and then adding a flag that indicates a null value when $N_{m,v}(i) = M(v)$. The quantum search routines that call $U_{N_{M, v}}$ can be easily modified to check for this flag.
For each vertex $v$, implementing $U_{N_{M, v}}$ in the above ways uses $O(\log |V|)$ overhead on top of the potential call to $U_{N_v}$ (see \cref{sec:qram}).

\paragraph{Quantum SCC finding.}
Existing work has produced quantum algorithms for determining if a graph is strongly connected~\cite{durr2006quantum}, noting that an adaptation of the approach can find the SCCs. 
In Appendix~\ref{sec:q-tarjan} we describe such an adaptation. We outline a quantum analog of Tarjan's algorithm, \textsc{Q-FindSCC}, which can output the SCCs of a directed graph with time complexity ${O}(\sqrt{nm} \log^2 n)$, when given quantum access to the graph, as just described (or $O(\sqrt{|V||E|} \log^2 |V|)$ in terms of the properties of $G$). The $\log$ factors come from requiring a constant success probability and low-level implementation details (including QRAM queries), as discussed in~\cref{sec:max-matching} for the maximum matching algorithms. In the next stage, we will need quantum access to the components $\mathcal{S}$, from a unitary $U_{\mathcal{S}}$; see Appendix~\ref{sec:q-tarjan}.
This can be done by recording them in QRAM during the execution of \textsc{Q-FindSCC} without changing the time complexity.

\paragraph{Finding edges.}
Finally, we describe how to remove edges with time complexity $O(\sqrt{|E||R|})$ with $\textsc{Q-IdentifyEdges}$, where $R$ is the set of edges that are removed. This procedure takes as input the set of unitaries $\{U_{N_v}\}_v$ and the unitary $U_{\mathcal{S}}$.
The general idea is to perform a quantum search over the $|E|$ edges in $G$. 
The time complexity of the procedure is ${O}(\sqrt{|E||R|}\log^2 |V|)$, even in the case when $|R|$ is unknown a priori~\cite{grover-bounds}.

Recall that we wish to remove edges that are: i) not in $M$, ii) not in $E_{\mathrm{used}}$, and iii) not connecting vertices in an SCC. We proceed by searching over the edges incident to variable vertices $v \in X$.
For each $v$ and $i\in [\delta_v]$, we want to remove the incident edge $\{v, N_v(i)\}$ iff
$ (\mathcal{S}(v) \neq \mathcal{S}(N_v(i)) ) \land (N_v(i) \neq M(v) ) \land (\{v,N_v(i)\} \notin E_{\mathrm{used}} )$.
Given quantum access to $\mathcal{S}, N_v$ and $E_{\mathrm{used}}$, this can be computed in $\log(|V|)$ time. 
Note, as with the matching, QRAM for quantum access to $E_{\mathrm{used}}$ can be initialized during the runtime of \textsc{FindSimplePaths} as edges are discovered, without adding to the complexity of \textsc{RemoveEdges}.

If there are $r_v \le \delta_v$ edges to be removed from the search over vertex $v$, the time complexity is $\tilde{O}(\sqrt{r_v \delta_v})$~\cite{grover-bounds, ambainis-q-search}.
Repeating this for each $v$ gives an aggregate time complexity of $O(\sqrt{|E||R|}\log^2 |V|)$ by the Cauchy-Schwarz inequality, using $\sum_v r_v=|R|$ and $\sum_v \delta_v=|E|$.

\paragraph{Analysis.} Combining the above with the classical \textsc{FindSimplePaths} gives a time complexity of $\tilde{O}\left( |E_{\mathrm{used}}| + \sqrt{|V||E|} + \sqrt{|E||R|} \right)$ for a quantum analog of \textsc{RemoveEdges}. In the worst case this is $\tilde{O}(|E|)$, matching the classical complexity (up to logarithmic factors). In cases where there are $|R|=O(|V|)$ edges to remove, and $|E_{\mathrm{used}}| = O(\sqrt{|V||E|})$, 
our quantum approach has time complexity $\tilde{O}(\sqrt{|V||E|})$, providing a $\sqrt{|E|/|V|}$ improvement (up to polylogarithmic factors) over the classical run time of $O(|E|)$.

%% file: sections/5-filtering/generalized.tex
\subsection{Generalizing quantum filtering}\label{sec:generalized}

As detailed in the previous section, filtering for the \texttt{alldifferent} constraint consists of a feasibility check followed by a pruning step to enforce domain consistency. The former is achieved by finding a maximum matching in a bipartite graph representation of the constraint, while the latter uses a combination of breadth-first and depth-first searches to look for SCCs to enable the pruning of edges from the graph. 
Other global constraints with a similar structure can also be accelerated using quantum subroutines. 
Indeed, the need to find matchings in graphs is a bottleneck for many global constraint filtering algorithms. The global cardinality constraint (\texttt{gcc}), for example, is another such constraint \cite{quimper2004improved}. 

\begin{definition}[\texttt{gcc} constraint] 
    Given variables $X = (x_1,\dots, x_{|X|})$, values $V = (v_1,\dots, v_{|V|})$, and cardinality bounds $\Gamma = (\gamma_1, \dots,\gamma_{|V|})$, where each $\gamma_i \in \Gamma$ is defined by an interval $[\ell_i, u_i]$, the constraint $\texttt{gcc}(X,V,\Gamma)$ requires that value $v_i$ take place in the solution between $\ell_i$ and $u_i$ times, inclusively.
\end{definition}

The \texttt{gcc} constraint is commonly used in scheduling, rostering, and timetabling problems~\cite{quimper2004improved}. Our previous work shows that the domain-consistency algorithm for \texttt{gcc} can be accelerated with quantum algorithms in a fashion similar to that for \texttt{alldifferent} \cite{booth2020quantum}. Beyond \texttt{alldifferent} and \texttt{gcc}, however, there are other global constraints whose domain-consistency algorithms consist of finding maximum matchings and SCCs in bipartite graphs, such as the $\texttt{alldifferent\_except\_0}$ constraint.

\begin{definition}[\texttt{alldifferent\_except\_0} constraint] 
Given variables $X = (x_1,\dots,x_{|X|})$, \texttt{alldifferent\_except\_0}$(x_1,\dots,x_{|X|})$ requires that all of the variables in its scope not assigned a value of $0$ to take on different values (i.e., in a solution to the constraint, $x_i \neq x_j, \forall i \neq j \in \{1,\dots,k\} \text{ where }  x_i \neq 0 \wedge x_j \neq 0$).
\end{definition}

In this case, the variable-value graph is constructed such that there are additional vertices on the value-side of the graph allowing multiple variables assigned to a value of $0$ to participate in a maximum matching and thus satisfy the constraint. This variation of \texttt{alldifferent} is often useful in models that require difference among a subset of the variables that is unknown \emph{a priori}. 

The work of Cymer~\cite{cymer2012dulmage,cymer2015gallai} provides a powerful mechanism for extending our results beyond \texttt{alldifferent} and \texttt{gcc}. 
Their initial work details a generic filtering algorithm incorporating the Dulmage-Mendelsohn (DM) canonical decomposition \cite{dulmage1958coverings} for global constraints whose domain-consistency algorithms consist of finding maximum matchings in bipartite graphs, followed by a subsequent linear-time step \cite{cymer2012dulmage}. 
Given a bipartite graph associated with the constraint, their general filtering mechanism has two main steps: finding a maximum matching and computing the DG canonical decomposition. 
As with R\'{e}gin's algorithm for \texttt{alldifferent}, their generic algorithm is dominated by the complexity of finding maximum matchings in bipartite graphs, as computing the DM decomposition can be done in linear time \cite{cymer2012dulmage}. 
It follows, then, that any global constraint that can be made domain consistent by their algorithm can also be accelerated, with respect to worst-case time complexity, by the quantum algorithm for finding maximum matchings, and the associated data structures, outlined in the previous section of this paper. 
In addition to \texttt{alldifferent}, \texttt{gcc}, and \texttt{alldifferent\_except\_0}, Cymer shows that their algorithm can be used for other global constraints including \texttt{inverse}, \texttt{same}, and \texttt{usedby}, as detailed in Appendix \ref{sec:other-globals}. 
See \cite{cymer2012dulmage} for the full list of thirteen global constraints considered.

In subsequent work \cite{cymer2015gallai}, Cymer proposes a generic filtering mechanism for global constraints whose domain-consistency algorithms are expressed over general graphs, incorporating the Gallai-Edmonds decomposition \cite{edmonds1965paths}. Similar to the bipartite case, maximum and optimal-degree matchings play an important role in this algorithm.  
Given that the quantum algorithm for finding maximum matchings in our approach is applicable to general graphs \cite{dorn2009quantum}, quantum filtering could accelerate aspects of the filtering for these families of constraints as well. 
In contrast to Cymer's DM-based algorithm, however, it is unclear as to whether quantum computing would improve on worst-case time complexity since finding matchings in their Gallai-Edmonds decomposition algorithm is not necessarily the dominating complexity term.

%% file: sections/6-backtracking-search.tex
\section{Quantum-accelerated branch-and-infer search}\label{sec:tree-search}

Recall that the constraint programming (CP) approach to solving a CSP is to augment backtracking search with logical inference. 
In the previous section, we showed how to accelerate inference with quantum processing. In this section, we focus on using quantum processing to accelerate  branch-and-infer search.
In~\cref{sec:classical-tree-search}, we detail how the quantum subroutines for inference described above can be integrated into an otherwise entirely classical backtracking search.
In~\cref{sec:quantum-tree-search}, we review existing quantum algorithms for backtracking search and 
introduce hybrid variants of quantum backtracking search that interpolate between the fully classical and fully quantum cases. 
Finally, in~\cref{sec:q-backtracking-with-inference}, we show how these quantum algorithms for backtracking can be applied to CP, including
how to integrate quantum filtering into quantum backtracking search algorithms to obtain a fully quantum branch-and-infer search algorithm.

\input{sections/6-backtracking-search/classical-backtracking}
\input{sections/6-backtracking-search/quantum-backtracking}
\input{sections/6-backtracking-search/quantum-backtracking-inference}

%% file: sections/6-backtracking-search/classical-backtracking.tex
\subsection{Classical backtracking with quantum-accelerated inference}\label{sec:classical-tree-search}

Integrating quantum filtering algorithms within a classical backtracking search is the most straightforward of the frameworks we propose, as well as potentially the earliest to be implementable on quantum devices with some degree of fault-tolerance since it requires the fewest quantum resources.
The high level idea is to use a classical processor to manage the tree search and a quantum co-processor to solve global constraint filtering subproblems. 
In the context of~\cref{alg:filter-domains}, this would involve replacing some (or all) of the $\textsc{Filter}_C$ algorithms with quantum analogs (i.e., using quantum $\texttt{alldifferent}$ filtering instead of the classical algorithm). 

Recall that our quantum subroutines require quantum access to their inputs. 
For simplicity, we assume that the contents of the QRAM can be prepared classically using the same low-level encoding before being loaded into the QRAM; in this way, any overhead from classical memory calls is exactly the same regardless of how the filtering is done. 
Quantum \textsc{FindMaximumMatching}, for example, requires quantum access to the bipartite variable-value graph~$G$, which we supposed is performed using QRAM, as discussed in \cref{sec:max-matching}. 
For the purposes of our classical backtracking approach with quantum-accelerated inference, it suffices to use a single QRAM capable of holding the bipartite variable-value graph for the most memory-intensive \texttt{alldifferent} constraint at the root of the tree (where it is largest). 
This QRAM can then be initialized from scratch for each filtering subproblem without impacting the asymptotic complexity of the quantum subroutine.\footnote{Recall that initializing the QRAM for bipartite variable-value graph $G$ takes time on the order of the number of edges in $G$.} 
We can also imagine a more sophisticated implementation where the QRAM is updated to reflect removed edges as the tree is traversed, instead of re-initializing it from scratch each time. 
We note both approaches result in the same asymptotic complexity. 

Similarly, the subroutine \textsc{Q-FindSCC} requires quantum access to the directed graph $G_M$. 
As discussed in~\cref{sec:remove-edges} the QRAM for $G$ (used by quantum \textsc{FindMaximumMatching}) can be converted to one for $G_M$ with negligible overhead.
The subroutine \textsc{Q-IdentifyEdges} requires quantum access to the strongly connected components $\mathcal S$ returned by \textsc{Q-FindSCC} and the used edges $E_{\mathrm{used}}$ returned by \textsc{FindSimplePaths}; the QRAM for these can be initialized during the course of these prior subroutines.
Again, only one QRAM is needed for each, and can be re-initialized for each iteration of filtering.

Given the probabilistic nature of the quantum algorithms we employ, their integration within the CP search must be done with care. 
Recall from~\cref{sec:max-matching} that each of the quantum subroutines proposed for $\texttt{alldifferent}$ filtering involves a polynomial number of quantum searches, where each instance of quantum search yields an incorrect answer with a probability bounded above by a constant; 
to ensure the overall success probability of the subroutines is bounded by a constant, each of the quantum searches is repeated $O(\log n)$ times. 
While this repetition yields a bounded-error algorithm for a single instance of filtering, in a tree search the filtering algorithms are called (potentially exponentially) many times, often more than once per node in the search. 
The repetitions necessary to ensure a constant overall success probability for an exponential number of quantum searches could  overwhelm any quantum speedups. 
With this in mind, we propose two approaches for integrating quantum filtering in a classical tree search: i) an exact method, where quantum subroutines with a certain property can be integrated without sacrificing the completeness of the search and ii) a bounded-error method.

\subsubsection{Exact method}

Quantum subroutines augmented to return an explicit failure indicator (e.g., a boolean flag that tells us whether or not the algorithm has succeeded) can be used in a tree search algorithm that requires perfect completeness (i.e., an exact tree search) with negligible overhead. To illustrate this, suppose we have such a quantum algorithm and a classical algorithm for the same problem (e.g., finding a maximum matching).
Let $c(n) = \poly(n)$ be the runtime of the classical algorithm, which always succeeds, and $q(n, p)$ be the runtime of the quantum algorithm with failure rate $p$, which we are free to choose.\footnote{Generally, if we have a quantum algorithm that succeeds with constant probability, we can get inverse polynomial failure probability with $O(\log n)$ repetitions, though for some algorithms such as quantum \textsc{FindMaximumMatching} this additional logarithmic factor is not needed.}
Suppose now that we run the quantum algorithm for $p = o(1 / c(n))$; if it fails (as noted by the failure indicator), we then run the classical algorithm.
This is a Las Vegas algorithm; it always succeeds, but its runtime varies probabilistically. 
The expected runtime is then $q(n, o(1 / c(n))) + p \cdot c(n) = q(n, 1 / \poly(n)) + o(1)$.
That is, the average runtime is only negligibly more than running the quantum algorithm for failure rate at most inverse polynomial.

Of course, the above result is only beneficial to us when the quantum algorithm offers a speedup over its classical counterpart. While we can often design quantum subroutines that offer such speedups, augmenting them to include the required failure indicator without impacting asymptotic runtimes is not always possible. To illustrate this point, consider the two subroutines in our quantum filtering algorithm for \texttt{alldifferent}, as detailed in \cref{sec:filtering}, in turn.

The quantum \textsc{FindMaximumMatching} subroutine can be extended to include the required failure indicator with the same complexity scaling reported in~\cref{sec:max-matching}. The failure rate of this subroutine, as previously stated, is already polynomially small, and a further reduction in failure rate to a smaller polynomial
entails only changing the constant
factor absorbed by the asymptotic notation.
To construct the required failure indicator, we must check whether or not the subroutine returned a maximum matching. In all cases, the subroutine as formulated returns some set of edges. In linear time, as we explain, we can check whether these edges are a maximum matching and set the failure indicator to ``true'' if not.
If the returned edges are a maximum matching, then in linear time we can construct a minimum vertex cover~\cite{bondy1977graph} of the same size.
If we follow this constructive procedure and the resulting vertex cover is larger than the matching, then we know it is not maximum and we set the failure indicator to ``true''.
Otherwise, we confidently return the maximum matching and a failure indicator of ``false''.

On the other hand, we cannot efficiently construct a failure indicator for the quantum \textsc{RemoveEdges} subroutine. It could return an incorrect edge to remove if \textsc{Q-FindSCC} returns an invalid set of SCCs. Unlike verifying a maximum matching, it is not possible to classically check whether the SCCs returned are correct without overwhelming the quantum speedup. As a result, this subroutine cannot be advantageously incorporated into an exact tree search approach.

Our quantum \texttt{alldifferent} filtering algorithm for an exact tree search implementation, then, would consist of both quantum and classical \textsc{FindMaximumMatching} subroutines and a classical \textsc{RemoveEdges} subroutine. 
A similar analysis can be conducted for the subroutines of other global constraint filtering algorithms. 

\subsubsection{Bounded-error and heuristic methods}

Alternatively, suppose we permit the overall tree search to fail (i.e., not find a solution to the CSP if one exists) with some constant probability, and we wish to use all quantum subroutines available to us to maximize the speedup achieved at each node. Since, in this case, the output of some quantum subroutines (i.e., quantum \textsc{RemoveEdges}) is not efficiently verifiable classically, we would would need an approach that is robust to errors.

To achieve a constant success probability for the overall tree search, we can restrict the search to calling our quantum subroutines a polynomial number of times; in this case it suffices to have each subroutine fail with probability at most inverse polynomial, which introduces at most $O(\log n)$ overhead, as discussed above.
We can also, for example, ensure that the quantum subroutines are invoked at earlier nodes in the tree which typically represent larger subproblems and will stand to benefit from quantum speedups the most. After some predetermined polynomially bounded threshold of calls to the quantum subroutines, the tree search transitions to using classical filtering only.

A final ``heuristic mode'' approach is to always use the quantum subroutines regardless of the size of the tree without stringent guarantees on the overall success probability. 
In this case, the effect of subroutine failures on the overall tree search is strongly dependent on the particular tree search and filtering algorithms used. 
For filtering algorithms in which the only failure mode is not pruning a domain value that could have been pruned, the tree search will remain complete. However, the resulting tree may end up larger than if the filtering succeeded without failure (i.e., pruned \emph{all} removable values).

%% file: sections/6-backtracking-search/quantum-backtracking.tex
\subsection{Quantum-accelerated backtracking}\label{sec:quantum-tree-search}
In this section we detail quantum-accelerated approaches to backtracking search. We begin by reviewing existing quantum backtracking search algorithms from the literature. Then, we present variations of these algorithms that perform quantum tree search over subtrees of the full tree, using fewer quantum resources at the expense of a smaller speedup. In \cref{sec:q-backtracking-with-inference} we will incorporate inference into the backtracking algorithms to obtain quantum branch-and-infer search algorithms.

Following existing work~\cite{montanaro2015quantum}, the results in this section assume the existence of a classical backtracking algorithm $\mathcal{A}$ that finds a solution to the CSP or determines that none exists. This algorithm implicitly defines a tree $\mathcal{T}$ that contains $T$ vertices and has depth $\depth$. As before, this classical backtracking algorithm is assumed to traverse the tree with a depth-first search, where the ordering of each node's children is determined by $\mathcal{A}$.
Further, we let $T_{\mathrm{UB}}$ and $L_{\mathrm{UB}}$ represent (efficiently calculable from the CSP itself) upper bounds on the number of nodes in and depth of $\mathcal{T}$, respectively. 
Finally, we let $T_{\mathcal{A}}$ be the number of nodes actually explored by $\mathcal{A}$ in finding a single solution to the CSP (or proving that none exists). 
We report the complexity of quantum backtracking algorithms as a function of these parameters; the full complexity includes the cost of implementing the per-node procedures, as detailed in~\cref{sec:q-backtracking-with-inference}.

\subsubsection{Background}\label{sec:generic-quantum-tree-search}

In quantum tree search, we are given quantum access to operators that locally define a tree (i.e., that specify each node's children) and a predicate that evaluates a node.
The goal can be i) to determine if the tree \emph{contains} a marked node, i.e., one for which the predicate value is 1 (indicating a solution to the CSP);
ii) to \emph{find} a marked node, if one exists; or iii) to find \emph{all} marked nodes. 
While the methods described here apply in a general setting, we will be concerned with their use to search the tree defined by a given backtracking procedure, as described in~\cref{sec:backtracking}.
We discuss implementation of the node operators in the context of CP in~\cref{sec:walk-op-implementation}.

Montanaro~\cite{montanaro2015quantum} (building off of Belovs~\cite{belovs2013quantum}) gave a quantum algorithm to determine if a search tree $\mathcal T$ contains a marked node in $\tilde{O}(\sqrt{T_{\mathrm{UB}} \depth_{\mathrm{UB}}})$ queries to the operators that locally define the tree, and to find such a node in $\tilde{O}(\sqrt{T \depth^3})$ queries (or $\tilde{O}(\sqrt{T \depth})$ when there's a promise of only one marked node).
Jarret and Wan (JW)~\cite{jarret2018improved} extended and improved the algorithm to find a marked node in $\tilde{O}(\sqrt{T \depth})$ queries in general. 
(Their query complexity is actually tighter when expressed in terms of the effective resistance of the tree, which is always upper bounded by the depth.)
Ambainis and Kokainis (AK)~\cite{ambainis2017quantum} gave a tree size estimation algorithm that, together with Montanaro's algorithm, yields an algorithm to find a marked node in $\tilde{O}(\sqrt{T_{\mathcal A} \depth^3})$ queries. That is, with a small overhead, AK ensures that the quantum backtracking algorithm explores only as much of the tree as the classical algorithm $\mathcal{A}$ would. $T_{\mathcal A}$, the actual number of nodes explored by $\mathcal{A}$, can be much smaller than $T_{\mathrm{UB}}$ for two reasons. 
First, the upper bound $T_{\mathrm{UB}}$ that we can efficiently calculate before starting the tree search may be much larger than the size $T$ of the tree, especially when $\mathcal{A}$ uses inference.
(Whether we have such a bound or not is irrelevant when actually finding a marked vertex, because we can try exponentially increasing values $T_{\mathrm{UB}}$, thereby introducing at most a factor logarithmic in $T_{\mathrm{UB}}$.)
Second, the classical algorithm $\mathcal A$ can stop when it finds the first solution, so the number of nodes $T_{\mathcal A}$ it explores can be much smaller than the total number of nodes $T$ in the tree when it 
is not required to find all solutions.
While AK's and JW's versions have better scaling than Montanaro's with respect to the number of nodes and depth, they involve more complicated procedures that can significantly increase the constant factors and degree of the logarithmic ones; a practical implementation would need to take this into account. Nevertheless, the quantum parts of all three variants are mostly the same, and so improvements in implementation for, say, Montanaro's algorithm would likely also apply to the the extensions.

The foundation of these quantum backtracking algorithms is a quantum walk operator defined as follows (see e.g.~\cite[Sec. 2]{montanaro2015quantum}).
For each vertex $s$ (with children $s'$) of the tree with root $r$ and any $\alpha >0$, let
\begin{equation}
\ket{\psi_s(\alpha)}
\propto
    \begin{cases}
    \ket{s} + \sqrt{\alpha} \sum_{s' \leftarrow s} \ket{s'}, & \text{if $s = r$},\\
    \ket{s} + \sum_{s' \leftarrow s} \ket{s'}, & \text{if $s \neq r$},
    \end{cases}
\end{equation}
where the summation $\sum_{s' \leftarrow s}$ is over all children of $s$.
Define a walk operator that reflects about the subspace perpendicular to $\ket{\psi_s(\alpha)}$ when $s$ is not marked:
\begin{equation}\label{eq:diffusion}
    \diffusion_s(\alpha) = \begin{cases}
    I, & \text{if $s$ is marked},\\
        I - 2 \ket{\psi_s(\alpha)} \bra{\psi_s(\alpha)}, & \text{otherwise},
    \end{cases}
\end{equation}
where $I$ is the identity operator.
Let $A$ be the set of vertices that are an even distance from the root (including the root itself), and $B$ be the set of vertices that are an odd distance from the root.
Let
\begin{equation}
\begin{split}
& \walk_A(\alpha) = \bigoplus_{s \in A} \diffusion_s(\alpha), \\
& \walk_B= \ket{r} \bra{r} +  \bigoplus_{s \in B} \diffusion_s(1).
\end{split}
\end{equation}
The quantum algorithms for tree search are based the spectral properties of the overall walk unitary 
$\walk_B \walk_A(\alpha)$.\footnote{While $W_A$ and $W_B$ have the form of quantum walk operators, the algorithms described here do \emph{not} use them to perform a quantum walk in the usual sense.}
Specifically, if there is at least one marked node, then the root $\ket{r}$ has non-trivial overlap with a $1$-eigenvector of $\walk_B \walk_A(\alpha)$; if there is no marked node, then the root is orthogonal to the $1$-eigenspace.
These can be distinguished by phase estimation on the root.

Montanaro's algorithm for detecting whether a marked node exists is to repeatedly perform quantum phase estimation for the operator $\walk_B \walk_A(\depth)$, 
starting with the initial state $\ket{r}$;
the algorithm returns affirmatively if enough of its eigenvalues are 1.
\footnote{As pointed out by~\cite{campbell2019applying}, because the algorithm only needs to distinguish between eigenvalue 1 and eigenvalues far from 1, the quantum Fourier transform at the end of the phase estimation can be replaced by a Hadamard on each qubit.}
Quantum phase estimation of a unitary operator $U$ to precision~$\delta$ uses $O(1/\delta)$ applications of controlled-$U$ and $O(\log(1/\delta))$ other gates~\cite{cleve1998quantum}.
To find a marked node, we can do classical descent on the tree, at each stage checking whether the subtree rooted at each child contains a marked vertex and descending on any one that does.
This multiplies the runtime by a factor of $\depth$.
Because we can check whether the output is actually a marked node, we can do the above for $T_{\mathrm{UB}}$ equal to increasing powers of 2, so that the overall runtime depends on the actual tree size $T$, up to polylogarithmic factors.
If there is a single marked node, then conditioned on the estimated phase being 1, the node register is a superposition of states each of which has half its amplitude on the root and the other half uniformly distributed over the path from the root to the marked vertex.
By measuring the node register and repeating the procedure on the subtree rooted at the output, we can get to the marked node in $O(\log \depth)$ repetitions.

JW's version proceeds similarly, except that phase estimation is run using $\walk_B \walk_A(\tilde{\eta})$, where $\tilde{\eta}$ is an estimate of the effective resistance of the tree\footnote{See~\cite{belovs2013quantum} for the definition of effective resistance of a graph and~\cite{jarret2018improved} for a more detailed discussion of its application.}.
The procedure to estimate $\tilde{\eta}$ consists of rounds of phase estimation (of the walk operator using the current estimate) followed by quantum amplitude estimation (of the 1-eigenspace), which has the same resource requirements as the phase estimation. By using the effective resistance, the overlap of the 1-eigenvector with the root is made to be about 1/2.
The state of the node register conditioned on the estimated phase being 1 is such that we can get to a marked node in the same way as Montanaro's procedure but in $O(\log \depth)$ repetitions for any number of marked nodes.

The algorithm~\cite{ambainis2017quantum} of Ambainis and Kokainis for quantum tree search works by searching the first $\tau$ nodes of the tree $\mathcal T$ that would be explored by a classical tree-search algorithm, for exponentially increasing values of $\tau$.
For each value of $\tau$, it does this by first generating a path $\mathbf u (\tau)= (u_0, \ldots, u_l)$ of length $l \leq \depth$ descending from the root $r=u_0$ that completely specifies the subtree $\mathcal T_{\tau}$ containing the first $\tau$ nodes;
the set of nodes of the subtree $\mathcal T_{\tau}$ specified by the path $\mathbf u(\tau)$ consists of the nodes of the path itself together with the nodes of each subtree rooted at any ``earlier'' sibling of any node in the path.
By earlier sibling of a node $s$, we mean a node $s'$ with the same parent as $s$ (i.e., a sibling) that is explored earlier by the classical algorithm.
An example is shown in~\cref{fig:first-chunk}, where the path consists of the right-most green colored nodes at each level.
Then Montanaro's algorithm is applied to the subtree $\mathcal T_{\tau}$ using a successor function modified according to $\mathbf u (\tau)$.
Recall that this entails performing phase estimation to precision $O(1 / \sqrt{\tau \depth})$, including for the final value of $\tau \approx T_{\mathcal A}$.

If we use JW's algorithm in place of Montanaro's in AK's algorithm in order to find a marked node within $T_{\mathcal A}$,
then the overall resource cost is $\tilde{O}(\sqrt{T_{\mathcal A} \depth^3})$ calls to the walk operator for the original tree $\mathcal T$ to get the path $\mathbf u$ and $\tilde{O}(\sqrt{T_{\mathcal A} \depth})$ calls to the walk operator for $\mathcal T_{\mathcal A}$ as specified by the path $\mathbf u$.
In~\cref{sec:walk-op-implementation}, we explain how to implement both walk operators for backtracking search schemes that integrate inference as in CP, but first we discuss partial quantum search variants of the above algorithms that require fewer quantum resources.

\subsubsection{Partially quantum tree search}\label{sec:hybrid-tree-search}

Resources will be constrained on quantum devices for years, likely decades, to come. Before being able to implement
the fully quantum tree search algorithms discussed in
\cref{sec:generic-quantum-tree-search}, it will be possible to implement methods that perform quantum search over smaller subtrees rather than over the full tree.  
Here, we describe two such methods. 

Rennela et al.\ also considered hybrid classical-quantum tree searches~\cite{rennela2020hybrid}. 
They were concerned with quantum algorithms bottlenecked by space constraints, and show how in certain cases divide-and-conquer algorithms can achieve genuine quantum speedups even with a number of qubits equal to some small fraction of the problem input size.
The essential idea is to apply the quantum algorithm to subtrees at the ``bottom'' of the search tree when the subproblems are sufficiently small.
In constrast, our hybrid algorithm allows us to apply the quantum tree search algorithm over subtrees that together cover the entire search tree, starting at the root. 
Unlike as with the divide-and-conquer approach, the effectiveness of our hybrid approach does not depend on the internal structure of the tree, but rather just its size and depth.
At a high level, our hybrid approach is motivated by the possibility that the limiting quantum resource is not space but accuracy of the computation.\footnote{While in theory this limitation applies in both the NISQ and early fault-tolerant regimes, we expect our algorithms to be effective only in the fault-tolerant regime, since they are unlikely to be successfully implementable using NISQ devices.
In the fault-tolerant regime,
accuracy can be increased using more \emph{physical} qubits, which supports the consideration of accuracy as a resource.}
Specifically, it may be possible that we can ``implement'' the operator $\walk_B \walk_A(\alpha)$, in the sense of having enough qubits to run the circuit, but that the noise level is such that phase estimation cannot be reliably performed to the precision $\tilde{O}(1/\sqrt{T})$ required to search the whole tree.

\paragraph{Chunky quantum tree search.}

\begin{figure}

\begin{subfigure}{\textwidth}
\begin{center}
\includegraphics[width=\textwidth]{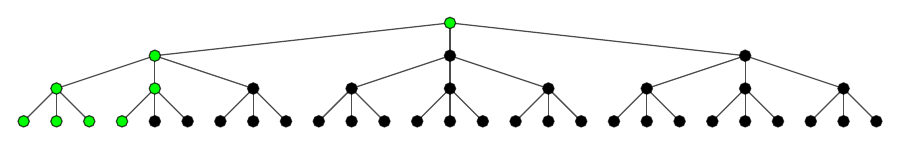}
\caption{
$\mathcal T_8 = \mathcal T_{0, 8}$
}\label{fig:first-chunk}
\end{center}
\end{subfigure}

\begin{subfigure}{\textwidth}
\begin{center}
\includegraphics[width=\textwidth]{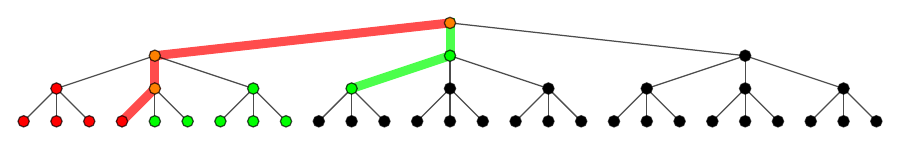}
\caption{$\mathcal T_{8, 16}$}
\end{center}
\end{subfigure}

\begin{subfigure}{\textwidth}
\begin{center}
\includegraphics[width=\textwidth]{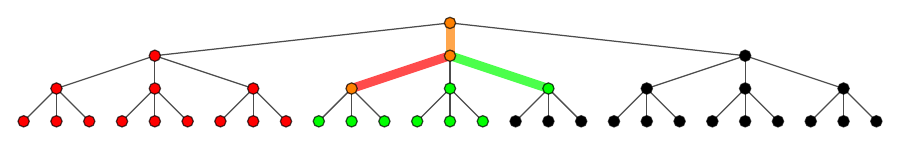}
\caption{$\mathcal T_{16, 24}$}
\end{center}
\end{subfigure}

\begin{subfigure}{\textwidth}
\begin{center}
\includegraphics[width=\textwidth]{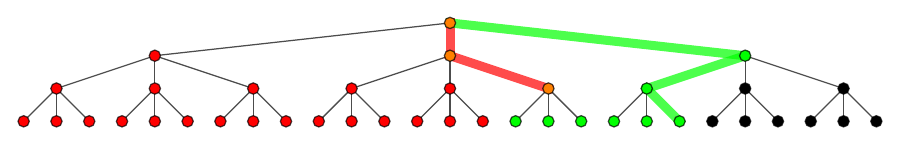}
\caption{$\mathcal T_{24, 32}$}
\end{center}
\end{subfigure}

\begin{subfigure}{\textwidth}
\begin{center}
\includegraphics[width=\textwidth]{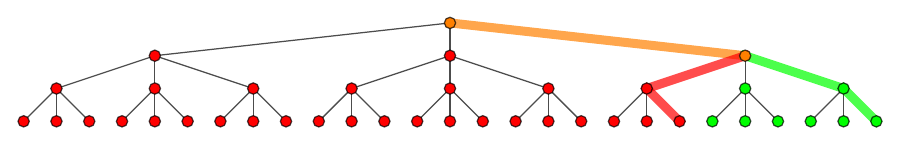}
\caption{$\mathcal T_{32, 40}$}
\end{center}
\end{subfigure}

\caption{
Visualization of chunky quantum tree search for chunks of size $\chi = 8$. 
For each $k=0, \ldots 4$, two subtrees are colored: $\mathcal T_{k \chi}$ and $\mathcal T_{k\chi, (k+1) \chi}$.
Nodes in $\mathcal T_{k \chi}$ but not $\mathcal T_{k\chi, (k+1)\chi}$ are red.
Nodes in $\mathcal T_{k \chi, (k+1)}$ but not $\mathcal T_{k\chi}$ are green.
Nodes in both are orange.
Similarly, two paths are shown.
Edges in the path defining $\mathcal T_{k \chi}$ but not in the path defining $\mathcal T_{k \chi, (k+1)\chi}$ are red.
Edges in the path defining $\mathcal T_{k \chi, (k+1)\chi}$ but not in the path defining $\mathcal T_{k \chi}$ are green.
Edges in both paths are orange.
}\label{fig:chunky-trees}

\end{figure}

We present now an extension of AK's algorithm with which the entire tree can be explored in arbitrarily sized chunks as illustrated in Figure~\ref{fig:chunky-trees}.
For simplicity, we assume all chunks are of uniform size $\chi$.
Let $\mathcal T_{\tau, \tau'}$ be the subtree (i.e., the ``chunk'') of $\mathcal T$ consisting of the first $\tau' - \tau$ nodes explored \emph{after} the first $\tau$ nodes, together with the union of the nodes on the unique path between the $\tau'$ nodes and the root.
Intuitively, $\mathcal T_{\tau, \tau'}$ is the difference between $\mathcal T_{\tau'}$ and $\mathcal T_{\tau}$, with minimal edges added to make it connected.
As discussed in~\cref{sec:generic-quantum-tree-search}, the subtree $\mathcal T_{\tau}$ can be uniquely specified by a path $\mathbf u(\tau)$. 
The subtree $\mathcal T_{\tau, \tau + \chi}$ can be uniquely specified by \emph{two} paths: $\mathbf u(\tau)$ and $\mathbf u(\tau + \chi)$.
The path $\mathbf u(\tau)$ can be thought of as defining $\mathcal T_{\tau}$ by specifying a youngest child for each node on the path.
The paths $\mathbf u(\tau)$ and $\mathbf u(\tau + \chi)$ can be thought of as defining $\mathcal T_{\tau, \tau + \chi}$ by additionally specifying an oldest and a youngest child for each node on their shared initial subpath.

Using AK's algorithm and for any $\tau$, we can get the path $\mathbf u(\tau)$ of length at most $\depth$ defining the subtree $\mathcal T_{\tau} = \mathcal T_{0, \tau}$ of the first $\tau$ nodes explored in $\mathcal{T}$.
Given paths $\mathbf u(\tau)$ and $\mathbf u(\tau + \chi)$, we can apply Montanaro's or JW's algorithm to explore the subtree $\mathcal T_{\tau, \tau + \chi}$ by modifying the walk operator. 

Each chunk $\mathcal T_{\tau, \tau + \chi}$ has at most $\chi + \depth$ nodes, and so can be explored using $\tilde{O}(\sqrt{(\chi + \depth)\depth})$ calls to the walk operator for $\mathcal T_{\tau, \tau + \chi}$.
Overall, the walk operator for $\mathcal T_{\tau, \tau + \chi}$ uses $\tilde{O}(\depth)$  more gates relative to the walk operator for $\mathcal T$, independent of $\tau$ and $\chi$.
Additionally, the path $\mathbf u(\tau + \chi)$ can be found using AK's algorithm using $\tilde{O}(\sqrt{(\chi + \depth)\depth^3})$ calls to the walk operator for $\mathcal T_{\tau, T_{\mathcal A}}$.

Assuming $\chi = \Omega(\depth)$, the specification and exploration of each chunk use $\tilde{O}(\sqrt{\chi \depth^3})$ calls to a modified walk operator.
There are $\ceil{T_{\mathcal A} / \chi}$ chunks, so overall there are $\tilde{O}(\sqrt{T_{\mathcal A} \depth^3/ \chi})$ calls.
For a single chunk ($\chi = T_{\mathcal A}$), we get the original runtime scaling of $\sqrt{T_{\mathcal A}}$, and for constant-sized chunks $\chi = O(1)$, we get the classical $T_{\mathcal A}$ scaling.
Importantly, the per-node cost is that of the largest cost node \emph{within each chunk}, which interpolates between the fully classical and fully quantum cases.

\paragraph{Bounded-depth quantum tree search.}

An alternative way of breaking up the tree into subtrees is to limit the quantum search by depth.
That is, starting at some node of the full search tree, we perform quantum search on the subtree rooted at that node and containing nodes at most some distance $\depth^{*}$ away.
This can be done by marking each node such that i) the predicate value is $1$ or ii) the predicate value is indeterminate and the node is distance $\depth^{*}$ from the root (of the subtree).
The outcome of each depth-$\depth^{*}$-bounded quantum search is either a satisfying solution if one exists within distance $\depth^{*}$ of the starting point, or the set of not-definitely-infeasible nodes at distance exactly $\depth^{*}$ away from the starting point.
The simplest case, $\depth^{*}=1$, is essentially Grover search over each node's children, which we can perform directly.
Specifically, if we classically know the node whose children we are searching over, we can classically compute 
its number of children and perform a modified Grover search to find all children $c$ that the predicate marks as not definitely infeasible.
Performing Grover search directly in this way allows us to store much of the state classically, while restricting our attention to children not known to be infeasible after propagation in a cheaper way.
More generally, such direct Grover searches are complicated by the generally unknown structure of the tree; see~\cite{furer2008solving,cerf2000nested}.

%% file: sections/6-backtracking-search/quantum-backtracking-inference.tex
\subsection{Quantum-accelerated backtracking with inference}\label{sec:q-backtracking-with-inference}

In this section we extend the ideas of the previous section to incorporate the propagation of CP. Classically, this involves as many rounds of filtering as needed to reach a fixed point (as exemplified in~\cref{alg:filter-domains}). Since, in the quantum case, the propagation must be done in superposition, the number (and sequence) of rounds of filtering must be fixed a priori. Similarly, the circuit for each filtering algorithm must be valid when applied to any corresponding value-variable graph throughout the tree; this means that the effective per-node filtering cost is that of the worst case. 

\subsubsection{Representation of tree nodes for CP}\label{sec:node-representation}

\input{sections/6-backtracking-search/node.tex}

Before getting to how to implement the walk operators, we will start by specifying one way of representing the nodes of the search tree in memory.
Concretely, we will represent each node of the search tree by  
the tuple
$
(
l,
\mathbf \branch,
\mathcal D,
\mathcal R
)$, where 
$l \in \{0, \dots, \depth\}$ is the number of branching constraints posted,
$\mathbf \branch \in {\left(\{\indeterminate\} \cup [\maxNumChildren]\right)}^{\depth}$ is the history of branching decisions (where $B$ is an upper bound on the number of children of any node in the tree), 
$\mathcal D = (D_1, \ldots, D_{|X|})$ denotes the domains (before filtering),
and 
$\mathcal R = (R_1, \ldots, R_{\depth})$ denotes the whole history of removed edges,
where $R_{l'}$ is the set of edges removed from the domains by the $l'$-th branching constraint.
\cref{tab:node} contains a more detailed description of how exactly this are represented at a low level, and how many qubits they require.

The main requirements of the representation of the search node are that each node is uniquely identified and that each node
contains all the information necessary get to that node's unique parent.
In general, different nodes in the search tree can have the same domains, so the domains alone are not enough.
The history $\mathbf \branch$ of branching decisions alone suffices in this sense, but we include the rest for ease of exposition and efficiency of computation.
In a slight abuse of notation, for two tuples $\mathcal A = (A_1, \ldots), \mathcal B = (B_1, \ldots) $ of sets (e.g. $\mathcal D$ or $\mathcal R$), we will write the difference as $\mathcal A \setminus \mathcal B = \bigcup_i A_i \setminus B_i$.  

The root of the full search tree has 
$l = 0$, 
$\mathbf b = (\indeterminate, \ldots)$,
$\mathcal D$ set to its initial value, 
and
$\mathcal R = (\emptyset, \ldots)$, but we will also consider searches of subtrees rooted at other nodes.
The children of a non-leaf node 
\begin{equation}
\left(l, \left(\ldots, b_l, \indeterminate, \ldots\right) \mathcal D, \left(\ldots, R_{l}, \emptyset, \ldots\right)
\right)
\end{equation}
are the nodes
\begin{equation}
{
\left\{
(l + 1,
\left(\ldots, b_l, c,
\ldots
\right)
,
\heuristic_c(\filter(\mathcal D))
,
\left(
\ldots, R_l, 
\mathcal D \setminus \heuristic_c(\filter(\mathcal D))
,
\ldots
\right)
\right\}}_{c \in [\numChildren(\mathcal D)]}
.
\end{equation}
Recall that $\filter(\mathcal D) \in 2^{\mathcal D}$ is the filtered domains and $\predicate(\mathcal D) \in \{\indeterminate, 0, 1\}$ indicates the known feasibility of the filtered domains.

This definition of the search tree is not unique, and several choices were made for ease of exposition.
Different representations of the nodes make different space-time tradeoffs.
That is, more consise representations can save space at the cost of requiring more computation.
A quantitative balance of these tradeoffs would depend on detailed knowledge of the hardware (including any underlying fault-tolerance scheme), of the choices of the heuristic and propagation functions, and of the class of instances.
Our definition also allows in the search tree nodes known to be infeasible (as does~\cite{campbell2019applying}), whereas in Montanaro's original formulation such a node was excluded when enumerating the children of its ostensible parent.
While including infeasible leaves in this way increases the number of nodes in the tree, it does not increase the number of times the filter operator (analogous to Montanaro's predicate) is called; in Montanaro's formulation, the predicate is called for each child in sequence, so that overall it is called a number of times proportional to the number of nodes in the inclusive tree.

\subsubsection{Implementation of walk operators for CP}\label{sec:walk-op-implementation}

Both operators $\walk_A(\alpha)$ and $\walk_B$ can be built from the following primitives:
the propagation operator
\begin{equation}\label{eq:filter-op}
U_F \ket{\mathcal D, \emptyset} \ket{\indeterminate}
=
\ket{\filter(\mathcal D), 
\mathcal D \setminus \filter(\mathcal D)
} \ket{\predicate(\mathcal D)}
,
\end{equation}
and
the branching operator
\begin{align}
&
U_{\mathrm{br},A}(\alpha)
\ket{0, 
\indeterminate,
\mathcal D, 
R
}
\nonumber
\\
&=
{\left(
1 + \numChildren \left(\mathcal D\right) \alpha
\right)}^{-1/2}
\left(
\ket{0, 
\indeterminate,
\mathcal D,
R}
+ 
\sqrt{\alpha}
\sum_{c=1}^{\numChildren(\mathcal D)}
\ket{1, 
c,
\heuristic_c(\mathcal D), 
R \cup \left(\mathcal D \setminus h_c(\mathcal D)\right)
}
\right)
,\label{eq:heuristic-op-0}
\\
&
U_{\mathrm{br}, A}(\alpha)
\ket{l, 
\indeterminate,
\mathcal D, 
R
}
\nonumber
\\
    &\overset{\text{even } l > 0}{=}
{\left(
1 + \numChildren \left(\mathcal D\right)
\right)}^{-1/2}
\Bigg(
\ket{l, 
\indeterminate,
\mathcal D, \mathcal R}
+ 
\sum_{c=1}^{\numChildren(\mathcal D)}
\ket{l + 1, 
c,
\heuristic_c(\mathcal D), 
R \cup \left(\mathcal D \setminus h_c(\mathcal D)\right)
}
\Bigg)
,
\label{eq:heuristic-op}
\end{align}
and $U_{\mathrm{br}, B}$ is defined as $U_{\mathrm{br}, A}$ for odd $l$.
By keeping track of the removed edges, we ensure that $U_F$ is invertible.
In other words, while additional ancillas may be needed to compute $U_F$, they can be returned to their initial state by the end.
If we have a classical circuit to compute $F$, then the quantum circuit for $U_F$ is approximately the same size.
In \cref{sec:q-filtering-and-tree-search}, we discuss how the quantum-accelerated version of the filter operator $U_F$ can be integrated into the quantum tree search.
Note that because we must perform the propagation in superposition, the cost per node is equal to the maximum over all nodes, rather than the average as in the classical case, where propagation at some nodes may be quicker than at others. This is the same whether we use a reversible version of a classical circuit, or a modification of quantum-accelerated filtering as described in~\cref{sec:q-filtering-and-tree-search}.
The implementation of the branching operator $U_{\mathrm{br}}$ will depend on the exact choice of the classical heuristic $\heuristic$, but below we give as an example a concrete implementation assuming classical circuits for $\numChildren$ and $\heuristic_{\child} (\cdot, c) = \heuristic_c (\cdot)$.

\begin{figure}
\caption*{
NB: For a generalized controlled gate $\ket{x} \ket{0} \mapsto \ket{x} U_x \ket{f(x)}$, we use a triangle to indicate the control register(s) $\ket{x}$; filled and unfilled circles have their usual meaning of binary controls and negated controls, respectively.
}
\begin{center}
\begin{subfigure}{\textwidth}
\begin{center}
\includegraphics[width=0.9\textwidth]{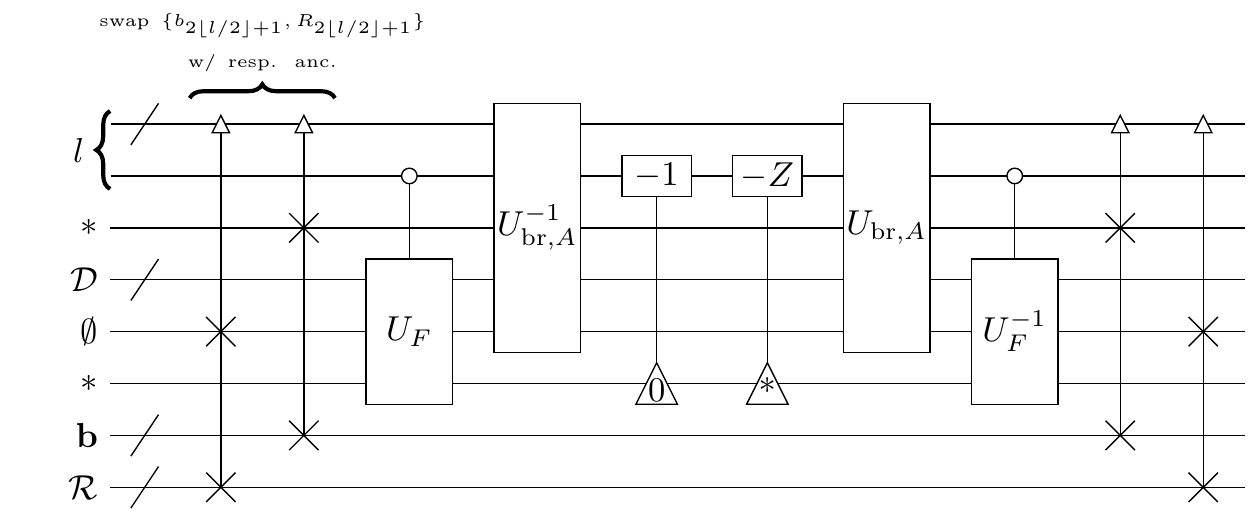}
\caption{
Circuit implementing $\walk_A(\alpha)$.
The bottom line of $l$ corresponds to the least significant bit indicating the parity of $l$.
The first gate swaps the $R_{2\lfloor l /2 \rfloor + 1}$ and the corresponding ancilla; the second does the same for $b_{2 \lfloor l/2\rfloor+1}$.
The central gates apply $-Z$ on the least-significant bit for predicate value $\flag = \indeterminate$ and a global sign flip for predicate value $\flag = 0$. 
}\label{fig:walker-circuit}
\end{center}
\end{subfigure}

\begin{subfigure}{\textwidth}
\begin{center}
\includegraphics[width=0.7\textwidth]{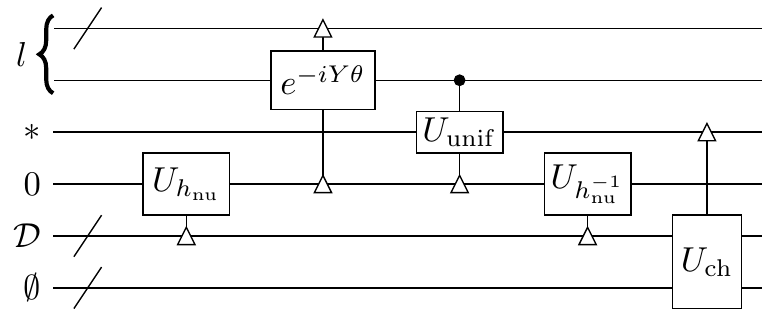}
\caption{
Circuit implementing $U_{\mathrm{br}, A}$.
In the second gate, $\theta = \tan^{-1} \sqrt{\alpha \numChildren}$ for $2\lfloor l / 2\rfloor = 0$ and $\theta = \tan^{-1} \sqrt{\ \numChildren}$ otherwise.
For example, when $2\lfloor l / 2\rfloor \neq 0$, the second gate takes $\ket{0}$ to (the normalization of) $\left(\ket{0} + \numChildren \ket{1}\right)$.
Circuit is only valid for even $l$.
}\label{fig:branching-circuit}
\end{center}
\end{subfigure}

\begin{subfigure}{\textwidth}
\begin{center}
\includegraphics[width=0.9\textwidth]{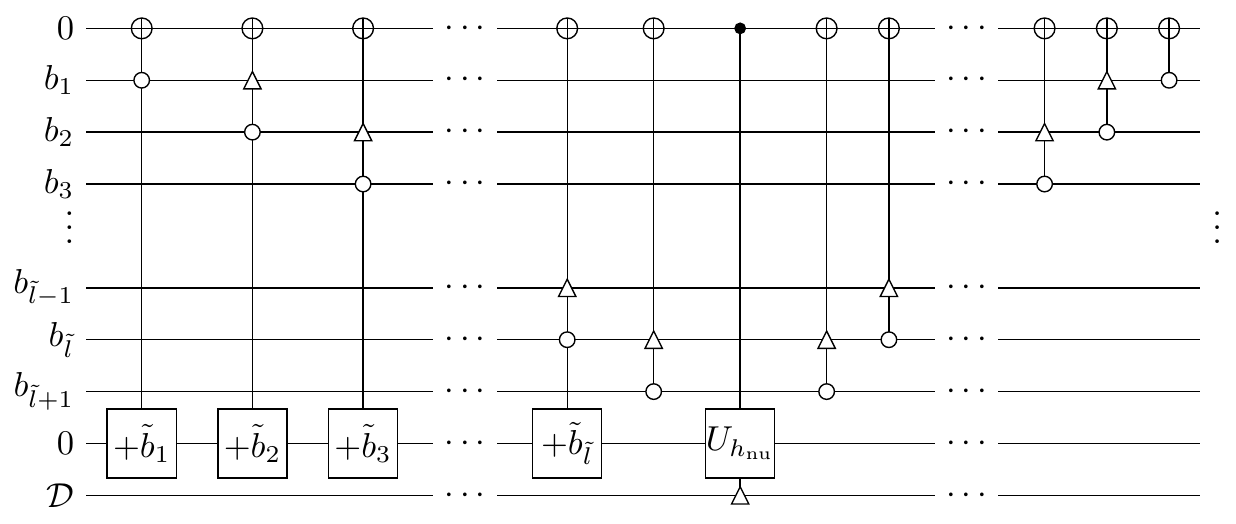}
\caption{
Circuit implementing $U_{\numEldestChildren}$.
For each $l \in [\tilde{l}]$ there is a gate that negates the first ancilla and adds $\tilde{\branch}_{l+1}$ to the second ancilla, conditioned on $(\branch_l, \branch_{l+1}) = (\tilde{\branch}_l, \indeterminate)$, with the convention that $\branch_{\tilde{l} + 1}=0$.
The first gate does the same but with the simpler condition that $\branch_1 = \indeterminate$.
The latter half of the gates reset the ancilla.
}\label{fig:mod-num-children-circuit}
\end{center}
\end{subfigure}

\caption{Circuits for walk operator and subroutines thereof.
Note that $2\lfloor l / 2\rfloor$ is obtained from $l$ by setting the least significant bit to zero.
}\label{fig:circuits}

\end{center}
\end{figure}

We will sketch how to implement $\walk_A(\alpha)$ out of these primitives; the implementation of $\walk_B$ is similar but simpler.
To start, note that $\walk_A(\alpha)$ acts on several subspaces independently, each spanned by a node at some even level $l$ and its children:

\begin{equation}
\ket{l, 
\left(
\ldots, b_l, \indeterminate, \ldots
\right),
\mathcal D, \left(
\ldots, R_l, \emptyset, \ldots
\right)}
\end{equation}
and
\begin{equation}
{\left\{
\ket{l + 1, 
\left(
\ldots, b_l, c, \ldots
\right),
\heuristic_c(\filter(\mathcal D)),
\left(
\ldots, R_l, 
\mathcal D \setminus
\heuristic_c(\filter(\mathcal D))
,\ldots
\right)
}
\right\}}_{c \in \left[ \numChildren(\filter(\mathcal D)) \right]}
.
\end{equation}
Suppose we add an ancilla state $\ket{\flag}$ for $\flag \in \{0, 1, \indeterminate\}$ and initialize it to $\ket{\indeterminate}$.
If we apply $U_F$ for even $l$, the children nodes are unchanged, and the parent node is transformed into
\begin{align}
&
\ket{l, 
\left(
\ldots, b_l, \indeterminate, \ldots
\right),
\filter(\mathcal D), \left(\ldots, R_l, \mathcal D \setminus \filter(\mathcal D), \ldots, \right)
}
.
\end{align}
Let $l_0$ be the least significant bit of $l$.
Now applying $U_{\mathrm{br}, A} (-Z_{l_0}) U_{\mathrm{br}, A}^{-1}$ reflects around a state similar to the right-hand sides of~\cref{eq:heuristic-op-0,eq:heuristic-op}, except with the parent filtered.
By controlling this on the predicate value $\flag = \indeterminate$, we ensure that the overall effect is the identity for marked vertices.
For predicate value $\flag = 0$, we apply a global sign flip (effecting~\cref{eq:diffusion} for an unmarked but childless node).
The circuit for this is shown in~\cref{fig:walker-circuit}.
A practical implementation of $\walk_A(\alpha)$ need not use $U_{\mathrm{br}, A}(\alpha)$ as a black box as above.

Now we explain how to implement the branching operator $U_{\mathrm{br}, A}$.
Assuming classical circuits for $\numChildren$ and $\heuristic_{\child} (\cdot, c) = \heuristic_c (\cdot)$, we can implement the operators
\begin{align}
U_{\numChildren} \ket{\mathcal D} \ket{0}
&
=
\ket{\mathcal D} \ket{\numChildren(\mathcal D)}
,
\\
U_{\child}
\ket{\indeterminate}
\ket{\mathcal D, R}
&=
\ket{\indeterminate}
\ket{\mathcal D, R}
,
\\
U_{\child}
\ket{c}
\ket{\mathcal D, R}
&=
\ket{c}
\ket{\heuristic_c(\mathcal D), R \cup \left(\mathcal D \setminus \heuristic_c(\mathcal D)\right)} 
.
\end{align}
We will also make use of a controlled uniform-superposition-producing operator
\begin{align}
U_{\uniform}
\ket{0} \ket{\indeterminate}
&= 
\ket{0} \ket{\indeterminate}
,
\\
U_{\uniform}
\ket{i} \ket{\indeterminate}
&\overset{i > 0}{=}
\ket{i}
\sum_{j=1}^i \ket{j}
.
\end{align}
Details on how to construct $U_{\uniform}$ can be found in~\cite[Sec. 4.7]{campbell2019applying}.
To implement the branching operator, we use an ancilla register, initialized to 0, to compute the branches produced by the heuristic.
(The case in which a node has no children because the predicate is true is accounted for in the circuit for $\walk_a(\alpha)$ \emph{outside} the subroutine $U_{\mathrm{br}, A}$, by cancelling out the branching operator with its inverse.)
Then conditioned on the number of branches $\numChildren$ and whether $l=0$, we produce a state proportional to $\ket{0} + \sqrt{\alpha \numChildren} \ket{1}$ on the least significant bit of $l$.
Conditioned on the least significant bit of $l$ being 1, we then prepare a uniform superposition over $[\numChildren]$ using $U_{\uniform}$.
We then uncompute $\numChildren$, and apply $h_c$ according to the ancilla.
The circuit for this is shown in~\cref{fig:branching-circuit}.
A practical implementation of $U_{\mathrm{br}, A}(\alpha)$ need not use $U_{\numChildren}, U_{\child}$ as black boxes as above.

AK's algorithm has two stages.
First, it generates the path $\mathbf u$ specifying the subtree $\mathcal T_{\tau}$ containing the first (by DFS) $\tau$ nodes of the full tree $\mathcal{T}$. 
It does this by repeated evaluations of a tree-size estimating procedure whose quantum part is just phase estimation of the walk operator $ \walk_B \walk_A(\depth)$ on successive nodes of the tree, with the slight modification that no marking is done (e.g., by conditioning the controlled phase in the middle of~\cref{fig:branching-circuit} on predicate values $\{0, 1\}$ rather than just $0$).
In our circuit for $\walk_A$ in~\cref{fig:walker-circuit}, this means just removing the control on the central $-Z$ gate.
Second, the algorithm performs Montanaro's (or JW's) algorithm for detecting or finding a marked node in the subtree $\mathcal T_{\tau}$ using the path $\mathbf u$.
In order to do that, we just need to modify the walk operator $\walk_B \walk_A (\alpha)$ so that it corresponds to $\mathcal T_{\tau}$ rather than the full tree $\mathcal T$.
We can do that as follows.
Let $\tilde{l} + 1$ be the length of $\mathbf u$.
For each node $u_l$ on the path, its values of $b_{l'}$ and $R_{l'}$ for all $l' \leq l$ are the same as those for nodes later in the path.
(The values of $\branch_{l'}$ and $R_{l'}$ in $u_l$ are $\indeterminate$ and $\emptyset$ for $l' > l$.)
Let $\tilde{\mathbf \branch}$ be the branching history for the last node $u_{\tilde{l}}$ in the path.
Define the function
\begin{align}
\numEldestChildren(l, \mathbf \branch, \mathcal D)
&=
\begin{cases}
0, & l = \tilde{l},\\
\tilde{\branch}_{l + 1}, & (\branch_1, \ldots, \branch_l) = (\tilde{\branch}_l, \ldots, \tilde{\branch}_l) ,\\
\numChildren(\mathcal D), & \text{o.w.,}
\end{cases}
\label{eq:mod-num-children}
\end{align}
which gives the number of branching decisions from a node in $\mathcal T_{\tau}$.
(We leave the dependence, via $\tilde{\mathbf \branch}$ and $\tilde{l}$, on the path $\mathbf u$ implicit, because we know that classically.)
Note that $\numEldestChildren$ depends on more parts of the state than $\numChildren$.
For a node not on the path, this is the same as for $\mathcal T$.
For a node on the path, this ensures that the next node on the path (if there is one) is the last branch in $\mathcal T_{\tau}$.
To implement Montanaro's algorithm on the subtree $\mathcal T_{\tau}$ defined by $\mathbf u$, it suffices to replace $U_{\numChildren}$ 
by 
\begin{equation}
U_{\numEldestChildren}
\ket{l, \mathbf \branch, \mathcal D}
\ket{0}
=
\ket{l, \mathbf \branch, \mathcal D}
\ket{\numEldestChildren(l, \mathbf \branch, \mathcal D)}
.
\end{equation}
We give an example circuit for this is in~\cref{fig:mod-num-children-circuit}.
Because $l$ is encoded in $\mathbf \branch$ (by the location of the last non-zero entry thereof), the circuit depends only on $(b_1, \ldots, b_{\tilde{l} + 1})$ and $\mathcal D$, in addition to some ancilla qubits.

\input{sections/6-backtracking-search/quantum-filtering}

%% file: sections/6-backtracking-search/node.tex
\begin{table*}
\begin{center}
\begin{tabular}{llll}
{\textbf{Variable}}                               &  \textbf{Range}                                & {\textbf{Role}}           &  {\textbf{Num. qubits}} \\
    {$l$}                                             &  $\{0, \ldots, \depth\}$                       & {num. of branching}  &  {$\ceil{\log(\depth + 1)}$} \\
                                                  &                                                &  {constraints posted}       \\
{$\mathbf \branch = (b_1, \ldots, b_{\depth})$}   &  ${\left(\{\indeterminate\} \cup [B]\right)}^{\depth}$   & {branches}                &  {$\depth \ceil{\log (\maxNumChildren + 1)}$} \\
$\mathcal D = (D_1, \ldots, D_{|X|})$             &  $2^{\mathcal D}$                              & ``active'' domains        &  {$\sum_{i=1}^{|X|} \left(\log (|D_i| + 1) + |D_i| \ceil{\log|V|}\right)$} \\
$\mathcal R = (R_1, \ldots, R_{\depth})$          &  ${\left(2^E\right)}^{\depth}$                 & edges $R_k$ removed       &  {$\depth (\log (m + 1) + m \ceil{\log |V|})$} \\
                                                  &                                                & after $k$th branching       
\end{tabular}
\caption{
Parts of our representation of a search node.
For the domains, the range and number of qubits is that of the original (largest) domains $\mathcal D$.
Each set $S$ (e.g., $D_i$ or $R_k$) is stored as an integer giving its size $|S|$ and a list $(s_1, \ldots, s_{|S|}, \nullassign, \ldots)$ of its entries followed by null values; the length of the list is the maximum size of the set.
The total number of qubits is $\tilde{O}(\depth m)$.
}\label{tab:node}
\end{center}
\end{table*}

%% file: sections/6-backtracking-search/quantum-filtering.tex
\subsubsection{Incorporating quantum algorithms for filtering}\label{sec:q-filtering-and-tree-search}

We now explain how the quantum-accelerated version of the filtering operator $U_F$ can be integrated into a fully or partially quantum tree search.
There are two main considerations: the QRAM needed as input and the possible error in the output.

We willl first address the QRAM.
Our quantum algorithms require quantum access to the variable-value graph $G = (X, V, E)$, as specified some active domains $\mathcal D$.
Here, we will work in the adjacency list model, but the situation is similar in the adjacency matrix model.
Let $N_{\mathcal D, v}$ be the neighbors of vertex $v$ in the variable-value graph implied by $\mathcal D$.
At the lowest level, each set $N_{\mathcal D, v}$ is represented as a list of its elements together with a register indicating its size.
For each $v$, there is an $\tilde{O}(|N_{\mathcal D, v}|)$-size, $\tilde{O}(\log|N_{\mathcal D, v}|)$-depth circuit implementing
\begin{equation}\label{eq:explicit-neighbor-query}
\ket{N_{\mathcal D, v}} \ket{i} \ket{0} \mapsto \ket{N_{\mathcal D, v}} \ket{i} \ket{N_{\mathcal D, v}(i)},
\end{equation}
i.e., an explicit circuit QRAM, as discussed in~\cref{sec:qram}.

At a given tree search node, we can initialize the QRAM by implementing
\begin{align}
\sum_{(l, \mathbf \branch, \mathcal D, \mathcal R)}
\psi_{l, \mathbf \branch, \mathcal D, \mathcal R}
\ket{l, \mathbf \branch, \mathcal D, \mathcal R} \bigotimes_{v \in V} \ket{\mathbf 0}
\mapsto
\sum_{(l, \mathbf \branch, \mathcal D, \mathcal R)}
\psi_{l, \mathbf \branch, \mathcal D, \mathcal R}
\ket{l, \mathbf \branch, \mathcal D, \mathcal R} \bigotimes_{v \in V} \ket{N_{\mathcal D, v}}
\end{align}
in $\tilde{O}(m)$ time, where the second ``QRAM'' register has $\tilde{O}(m)$ qubits.
The contents of the QRAM are then in superposition, but entangled with the node register.
The filtering algorithm can then query the graph (perhaps for a superposition over query registers $\ket{i}$) by calling the operator in~\cref{eq:explicit-neighbor-query}.
There are two potential sources for further efficiency.
First, each filtering algorithm only needs access to the subgraph induced by the variables in the corresponding constraint's scope and their neighbors.
Second, as discussed in~\cref{sec:node-representation}, the part of the node register containing $\mathcal D$ is exactly the registers $\ket{N_{\mathcal D, v}}$ for variable vertices, so there is no need to copy them into a separate ancilla register.

Now we will address the error in the output.
In~\cref{sec:filtering}, we described the quantum filtering algorithm as a mostly classical procedure that utilizes quantum search, each time completely measuring the quantum state before proceeding.
However, as described in~\cref{sec:unitarization}, we can construct a single unitary that performs the filtering without changing the error or runtime.
Unlike as when doing classical backtracking, we cannot amortize away the classical failsafe checks, and so there will always be some probability of failure.
Our choices then are to only use the accelerated filtering in a polynomial number of places, or to use it everywhere without any assurance that the overall tree search will be successful.
We also need to formulate the filtering algorithm such that the final state (when the algorithm is successful) is independent of the intermediate measurement results.
Whenever the filtering algorithm succeeds, it outputs the same set of edges to remove.
If the filtering is part of a classical procedure, the order of the elements in this set would not matter.
However, within the classical algorithm, we need the state $\ket{\mathcal D', R, \predicate(\mathcal D)}$ to be uniquely specified at the lowest level; this can be easily achieved, for example, by storing the domains $\mathcal D'$ and the removed edges $R$ as sorted lists.

%% file: sections/7-conclusions.tex
\section{Conclusions} \label{sec:conclusions}

In this paper, we investigate the use of quantum computing to accelerate constraint programming (CP).
We adapt recent work in quantum algorithms for graph problems and propose quantum subroutines to accelerate the filtering of the \texttt{alldifferent} constraint and other global constraints whose domain-consistency filtering problems involve finding a maximum matching in a bipartite graph. We detail frameworks for integrating quantum-accelerated inference algorithms within classical and quantum backtracking search schemes. Our work highlights the potential for mutual benefit between the paradigms of quantum computing and CP. 

One avenue for future work is to investigate whether even faster quantum algorithms for constraint filtering can be obtained by leveraging recent advances in quantum query complexity for various graph problems \cite{lin2014upper,beigi2020quantum,kimmel2020query}. 
Such query complexity improvements suggest that further quantum speedups may be possible in terms of time complexity. 
Similarly, advances in quantum algorithms more generally may allow speedups to be shown for a wider variety of constraint types. 
A particularly promising direction for future research is the use of quantum computers to accelerate domain-consistency algorithms posed in terms of a Gallai-Edmonds decomposition \cite{cymer2015gallai} as noted in \cref{sec:generalized}.

%% file: sections/8-acknowledgements.tex
\section*{Acknowledgements}
The authors are grateful for the support of the NASA Ames Research Center.
K.B., J.M., and S.H. were supported by NASA Academic Mission Services (NAMS), contract number NNA16BD14C. K.B. was also supported by the NASA Advanced Exploration Systems (AES) program. B.O. was supported by a NASA Space Technology Research Fellowship and the NSF QLCI program through grant number OMA-2016245.

Quantum circuit diagrams were made using $\langle\mathsf{q}|\mathsf{pic}\rangle$~\cite{qpic2020qpic}.

%% file: sections/appendix/quantum_tarjan.tex
\section{Quantum algorithm for strongly connected components}\label{sec:q-tarjan}

Here we propose a quantum version of Tarjan's algorithm to find the strongly connected components (SCCs) in a directed graph $G = (V,E)$.
As mentioned in the main text, \Durr\ et al.~\cite{durr2006quantum} say that their algorithm for determining strong connectivity can be extended to find the components; here, we give such an extension explicitly.
Our version closely follows the flow of the classical Tarjan algorithm, which uses bookkeeping during a DFS to find the SCCs~\cite{tarjan}.
We replace two of the classical searches by quantum searches to achieve a speedup from $O(m)$ to $O(\sqrt{nm} \log^2 n)$.

The procedure is detailed in~\cref{alg:q-tarjan,alg:q-sc}.
The essential component is \textsc{Q-StrongConnect}, which is called recursively to perform the DFS, recording the SCCs as they are found.
As in the standard Tarjan algorithm, several classical data structures are maintained:
\begin{itemize}
\item
A stack $S$ of vertices, to which each vertex is pushed upon discovery and from which it is removed upon assignment to an SCC.\@
We will require quantum access to the presence or absence of a vertex on the stack,
\begin{equation}\label{eq:U_S}
U_S \ket{v} \ket{0} \mapsto \ket{v} \ket{S(v)},
\end{equation}
where $S(v) = 1$ if $v$ is on the stack and $S(v) = 0$ otherwise.
\item
For each vertex $v$, an index $\dfsindex(v) \in [|V|] \cup \{\varnothing\}$, where $\varnothing$ is the initial null value, indicating the order in which $v$ was discovered by the DFS.
We will require quantum access to $\dfsindex(v)$:
\begin{equation}\label{eq:U_index}
    U_{\dfsindex} \ket{v} \ket{0}
    \mapsto
    \ket{v} \ket{\dfsindex(v)}.
\end{equation}
\item
For each vertex $v$, a ``low link'' $\lowlink(v)$ indicating the smallest index of a vertex reachable from $v$, including itself.
\end{itemize}

The two quantum subroutines, \textsc{Q-FindUndiscoveredNeighbor} and \\ \textsc{Q-FindMinIndexNeighborOnStack}, each take as input a vertex, and perform a search over its forward neighbors in time $\tilde{O}(\sqrt{\degree_v})$ where $\degree_v$ is the degree of the vertex $v$.
They require access to an oracle for labelling discovered vertices and an oracle for outputting the index value, respectively. 

\textsc{Q-FindUndiscoveredNeighbor}($v$, $U_{\ngbrs_v}$, $U_{\dfsindex}$) searches over $i \in [\degree_v]$ for an $i$ such that $\dfsindex(\ngbrs_v(i)) = \varnothing$ and returns the actual vertex $\ngbrs_v(i)$.

\textsc{Q-FindMinIndexNeighborOnStack}($v$, $U_{\ngbrs_v}$, $U_{\dfsindex}$, $U_S$)
attempts to find
\begin{equation*}
\argmin_{i: S(\ngbrs_v(i)) = 1} \{\dfsindex(\ngbrs_v(i))\}
\end{equation*}

This can be achieved in $\tilde{O}(\sqrt{\degree_v})$ time using a slight modification of an algorithm for minimum finding due to {\Durr}  and \Hoyer~\cite{q-find-min}.
The algorithm consists of repeated Grover searches using the predicate
\begin{equation}
(\dfsindex(\ngbrs_v(i)) < \dfsindex(\ngbrs_v(i^*))) \land S(\ngbrs_v(i)) = 1,
\end{equation}
where $i^*$ is initially set uniformly at random from $[\degree_v]$ and reset to the outcome of every successful Grover search.
The procedure finds the minimum with constant probability, in $\tilde{O}(\sqrt{\degree_v})$ time, including queries to $U_{\ngbrs_v}$, $U_S$, and $U_{\dfsindex}$.

\textsc{Q-FindUndiscoveredNeighbor} and \textsc{Q-FindMinIndexNeighborOnStack} are called $O(n)$ times (to traverse all $n$ vertices in the DFS).
To get a constant error probability for \textsc{Q-FindSCC}, each should be repeated $\log n$ times to get their error rates to $O(1/n)$.
Overall, the running time is $\tilde{O}\left(\sum_{v \in V} \sqrt{\degree_v}\right) = \tilde{O}(\sqrt{n m})$. This assumes that $ U_{\ngbrs_v}$ for all $v \in V$ is previously initialized. 
$U_S$ and $U_{\dfsindex}$ can be initialized in $\tilde{O}(n)$ 
and updated in $O(\log n)$.

\begin{algorithm}
\SetAlgoLined
\KwData{Directed graph $G$ with quantum access $\mathcal U_{\ngbrs} = {\left(U_{\ngbrs_v}\right)}_{v \in V}$.} 
\KwResult{Strongly connected components $\mathcal{S} = {\left(\mathcal{S}_v\right)}_{v \in V}$.}
\For{$v \in V$}{
$\dfsindex(v) \leftarrow \varnothing$ \tcp{quantum} 
$\lowlink(v) \leftarrow \varnothing $\;
}
index $\leftarrow$ 0\;
$S \leftarrow$ empty stack \tcp{quantum}
$\mathcal{S} \leftarrow {\left(\varnothing\right)}_{v \in V}$\;
\For{$v\in V$}{
\If{$\dfsindex(v)$ is $\varnothing $}{
$\textsc{Q-StrongConnect}(v, \text{index})$}
}
\caption{\textsc{Q-FindSCC}}
\label{alg:q-tarjan}
\end{algorithm}

\begin{algorithm}
\SetAlgoLined
\KwData{Vertex $v$ and index (passed by reference)} 
\KwResult{Updated component values $\mathcal{S}$}
 $\dfsindex(v) \leftarrow$ index\; 
 $\lowlink(v) \leftarrow$ index\;
index $\leftarrow$ index+1\;
push($S$, $v$)\;  
\While(){True}{
$w \leftarrow $ \textsc{Q-FindUndiscoveredNeighbor}($v$,
    $U_{\ngbrs_v}$, $U_{\dfsindex}$
    )\; 
\uIf{$\dfsindex(w)$ is $\varnothing$}{
$\textsc{Q-StrongConnect}(w, \text{index})$\;
$\lowlink(v)$ $\leftarrow$ min(\lowlink(v), \lowlink(w))\; 
}
\Else{break}
}

    $w \leftarrow $ \textsc{Q-FindMinIndexNeighborOnStack}($v$, $U_{\ngbrs_v}$, $U_S$)\;
$\lowlink(v)$ $\leftarrow$ min($\lowlink(v)$, $\dfsindex(w)$)\; 

 \If{$\lowlink(v) = \dfsindex(v)$}{
\While{True}{
    $u$ $\leftarrow$ pop($S$)\; 
    $\mathcal{S}[u] \leftarrow \lowlink(v)$\;

\If{$u=v$}{
break\;}
}
}
\caption{\textsc{Q-StrongConnect}}
\label{alg:q-sc}
\end{algorithm}

%% file: sections/appendix/other_globals.tex
\newpage
\section{Other global constraints}\label{sec:other-globals}

Additional global constraints can be made domain consistent using the Dulmage-Mendelsohn canonical decomposition algorithm of Cymer, and thus can be accelerated by our quantum approach as discussed in \cref{sec:generalized}. Here we consider several examples; 
see \cite{cymer2012dulmage} for a list of thirteen applicable global constraints. 

\begin{definition}
[\texttt{inverse} constraint] Let $N = \{1,\dots,n\}$. Given tuples of variables $X = (x_1, \dots, x_n)$ and $Y = (y_1,\dots,y_n)$ with domains $D_{x_i} = D_{y_i} = N, \forall i \in N$, the constraint $\texttt{inverse}(X,Y)$ enforces that: $x_{i} = j \Leftrightarrow y_{j} = i \text{ for all pairs } i,j \in N$ such that $i \neq j$.  
\end{definition}

The \texttt{inverse} constraint is often used to model routing problems involving successor and predecessor variables. For example, a CP model of the symmetric traveling salesman problem (TSP) with $n$ vertices can be posed as follows:
\begin{align}
\min \ & \sum_{i \in N} c_{i,x_i}   \\
\text{s.t}. \ & \texttt{inverse}(X, Y) \\
 & \texttt{alldifferent}(X) \\ 
 & \texttt{alldifferent}(Y) \\ 
 & x_1 < y_1 \label{symmetry-breaking}  \\ 
& x_i \in N, y_i \in N & \forall i \in N
\end{align}

\noindent where $c_{ij}$ is the cost of traveling from vertex $i$ to $j$, $x_i$ is the vertex visited immediately after $i$ and $y_i$ is the vertex visited immediately before $i$. Constraint \eqref{symmetry-breaking} breaks symmetry by preventing reverse tours~\cite{benchimol2012improved}. A solution that visits the vertices in order $(1,2,3,4)$ would have $X = (2,3,4,1)$ and $Y = (4,1,2,3)$. 

\begin{definition}
[\texttt{same} constraint] Given tuples of variables $X= (x_1,\dots,x_n)$ and $Y=(y_1,\dots,y_n)$, the constraint $\texttt{same}(X,Y)$ enforces that the multiset of values assigned to the variables of $X$ is identical to the multiset of values assigned to the variables of $Y$ \cite{beldiceanu2004filtering}.  
\end{definition}

For example, $X = (1,1,2,4)$ and  $Y=(1,2,1,4)$  would satisfy the constraint $\texttt{same}(X,Y)$, whereas $X = (1,2)$, $Y=(1,1)$ would not. If the cardinality of the two sets is not equal (i.e., $|X| > |Y|$), the $\texttt{usedby}$ constraint can instead be used.

\begin{definition}
[\texttt{usedby} constraint] Given tuples of variables $X = (x_1,\dots,x_n)$ and $Y=(y_1,\dots,y_m)$, where $m \leq n$, the constraint $\texttt{usedby}(X,Y)$ enforces that the multiset of values assigned to the variables of $Y$ is contained in the multiset of values assigned to the variables of $X$ \cite{beldiceanu2004filtering}.  
\end{definition}

For example, $X=(1,2,3)$ and $Y=(2,1)$ would satisfy the constraint $\texttt{usedby}(X,Y)$, whereas $X=(1,2,3)$ and $Y=(4)$ would not.  The \texttt{usedby} constraint is often useful for models that require an assignment of resources where there may be more of one resource type than another (i.e., nurses and doctors). 

%% file: sections/appendix/models.tex
\section{Other CP modeling examples}\label{sec:other-models}

In this appendix we provide additional examples to illustrate how computationally difficult problems can be modeled in constraint programming (CP). These examples are on top of the Sudoku model presented in \cref{sec:background} and the TSP model presented in \cref{sec:other-globals}, and utilize only the global constraints formally defined in this paper. 

\subsection{Rostering problems}

CP has been used to model and solve various scheduling problems, such as the nurse rostering problem~\cite{qu2008hybrid} from operations research. In this problem we are given a team of nurses, $i \in I$; a set of days in the rostering period, $d \in D$; a set of different shifts, $k \in K = \{M, A, E, O\}$ (i.e., morning, afternoon, evening, and off); and a pre-specified rostering requirement, $C_{d,k}$, specifying the number of nurses required on day $d$, shift $k$. We can model this rostering requirement in CP effectively using the global cardinality constraint ($\texttt{gcc}$), as introduced in \cref{sec:generalized}.

We define decision variable $s_{i,d} \in K$ to represent the shift assigned to nurse $i$ on day $d$. Our $\texttt{gcc}$ constraint is then formulated as follows:

\begin{equation}
    \texttt{gcc}(\{s_{i,d} : i \in I\}, \{M, A, E\}, \{C_{d,M}, C_{d,A}, C_{d,E}\}), \forall d \in D. \label{nurse-gcc}
\end{equation}

\noindent Constraint~\eqref{nurse-gcc} requires that, for example, there are exactly $C_{d,M}$ morning shifts assigned to the team of nurses on day $d$. It is also common to express that each nurse receive between $W_{\min}$ and $W_{\max}$ days off in the schedule. This can be accomplished with the following $\texttt{gcc}$ constraint:

\begin{equation}
    \texttt{gcc}(\{s_{i,d} : d \in D\}, \{O\}, \{[W_{\min}, W_{\max}]\}), \forall i \in I.
\end{equation}

\subsection{Sports tournament scheduling}

Another interesting application of CP modeling is for round-robin sports tournament scheduling~\cite{trick2002integer}. In a single round-robin tournament we have a set of $n$ teams, $i \in I$, and a set of $n-1$ match slots, $t \in T$. The goal is to design a schedule such that each team plays each other team exactly once. (There are also more complex versions of round-robin tournament scheduling such as bipartite single round robin~\cite{trick2002integer}.) 

To formulate this problem in CP, we introduce decision variable $x_{i,t} \in I \setminus \{i\}$ whose value indicates the opponent team $i$ faces in match slot $t$. The constraints can then be expressed as:
\begin{align}
    & x_{x_{i,t},t} = i, & \forall t \in T \label{sports-1} \\
    & \texttt{alldifferent}(\{x_{i,t} : t \in T\}), & \forall i \in I \label{sports-2} \\ 
    & x_{i,t} \in I \setminus \{i\}, & \forall i \in I, t \in T. \label{sports-3}
\end{align}

\noindent Constraint~\eqref{sports-1} ensures that the team that $i$ plays in slot $t$ plays $i$ in slot $t$; the flexibility of the CP paradigm allows variables to be indexed by other variables.\footnote{Some solvers require this to be modeled using the \texttt{element} constraint~\cite{nethercote2007minizinc}, while others allow this indexing to be directly expressed.} Constraint~\eqref{sports-2} dictates that each team has a different opponent in each match slot, and finally Constraint~\eqref{sports-3} dictates the domain for each of the variables.

While this modeling example illustrates the core constraints for round-robin scheduling, real-world models typically capture more complex constraints, such as those surrounding limitations on the number of `away' games, or the incorporation of travel times.  

\subsection{Quadratic assignment problems}

While the previous examples were both satisfaction problems, CP can be used to model and solve optimization problems as well, including those with quadratic objective functions. Consider a problem where a set of $n$ facilities, $I$, must be assigned to a set of $n$ different locations, $L$, and the distance between locations $\ell$ and $k$ is given by $d_{\ell,k}$. Each pair of facilities $(i,j)$ is also associated with a weight, $w_{i,j}$. We seek to minimize the weighted distances.

We can model this in CP by first introducing the integer decision variable $x_i \in L$ whose value represents the location that facility $i$ is assigned to. The objective function is then stated as follows:

\begin{equation}
    \min \sum_{i \in I, j \in I, i \neq j} d_{x_i,x_j} w_{i,j}
    .
\end{equation}

In this case, we use decision variables $x_i$ to index the distance parameter matrix $d_{\ell,k}$. We can enforce that the location of each facility must be different by adding the constraint $\texttt{alldifferent}(\{x_i : i \in I\})$ to the model.

%% file: sections/appendix/unitarizing_quantum_filtering.tex
\section{Wrapping quantum subroutines into a single unitary}\label{sec:unitarization}

Here we detail how to take an algorithm that consists of classical computations interspersed with quantum subroutines and produce a single unitary with the same effect, error probability, and runtime scaling.

Suppose we want to compute a bijective function $f:\Sigma \rightarrow \Gamma$. 
Consider a classical stochastic algorithm $\mathcal A$ that computes the function $f(\boldsymbol \sigma)$ in $k$ steps, sampling from a random variable $\outcomeSet_l$ before each step $l$.
Let $\boldsymbol \sigma = \boldsymbol \sigma_0$, $\boldsymbol \sigma_1$, \ldots, $\boldsymbol \sigma_k \in \Sigma$ be the intermediate states of the algorithm, and let $f_l(\boldsymbol \sigma_{l}, \outcome_l) = \boldsymbol \sigma_{l+1}$ be the deterministic function that advances the state.
Each intermediate state $\boldsymbol \sigma_l$ is stochastic, but completely determined by the previous measurements $\outcomes^{(l)} = (\outcome_0, \outcome_1, \ldots \outcome_{l-1})$, and so we write 
$\boldsymbol \sigma_l(\outcomes^{(l)})$.

Assume that each step $f_l$ is reversible, i.e. $(\boldsymbol \sigma_l, \outcome_l)$ is completely determined by $\boldsymbol \sigma_{l+1}$.
Then a classical circuit description of $f_l$ can be translated into a quantum circuit that computes the transformation
\begin{equation}
U_{f_l} \ket{\boldsymbol \sigma_l, \outcome_l} = \ket{\boldsymbol \sigma_{l+1}}
\end{equation}
in the usual way.
Now, suppose that we can implement
\begin{equation}\label{eq:U_M}
U_{\outcomeSet_l} \ket{\boldsymbol \sigma_l, \mathbf 0}
=
\sum_{\outcome_l \in \outcomeSet_l(\boldsymbol \sigma_l)}
\sqrt{\Pr[\outcome_l | \boldsymbol \sigma_l]}
\ket{\boldsymbol \sigma_l, \outcome_l},
\end{equation}
which computes a quantum state corresponding to the probability distribution $\Pr[\outcomeSet_l | \boldsymbol \sigma_l]$.
Then together we can compute
\begin{align}\label{eq:UfUM}
&U_{f_{k-1}}
U_{\outcomeSet_{k-1}}
\cdots
U_{f_0}
U_{\outcomeSet_0}
\ket{\boldsymbol \sigma_0} 
\nonumber\\
&=
\sum_{\mathclap{\outcomes^{(k)} \in \outcomeSets^{(k)}}}
\sqrt{\Pr[\outcomes^{(k)}]}
\ket{\boldsymbol \sigma_k (\outcomes^{(k)})}
\end{align}
where we introduced the notation $\outcomeSets^{(l)} = (\outcomeSet_0, \outcomeSet_1, \ldots, \outcomeSet_{l-1})$ and $\outcome \in \outcomeSet$ means that $\outcome$ is a possible value of the random variable $\outcomeSet$.

Lastly, by construction we can write the final state as 
$\boldsymbol \sigma_k(\outcomes^{(k)}) = (f(\boldsymbol \sigma), \gamma(\outcomes^{(k)}))$, i.e., as the concatenation of the answer and some other information $\gamma(\outcomes^{(k)})$, including that which enables reversibility.
Then we can perform the transformation in~\cref{eq:UfUM}, copy $f(\boldsymbol \sigma)$ into an ancilla register, and then uncompute to effect
$\ket{\boldsymbol \sigma_0}  \ket{\mathbf 0} \mapsto \ket{\boldsymbol \sigma_0} \ket{f(\boldsymbol \sigma_0)}$.
If we additionally can compute $f^{-1}$, then overall we can implement
\begin{equation}
U_f \ket{\boldsymbol \sigma_0} = \ket{f(\boldsymbol \sigma_0)}.
\end{equation}

The key idea is that, while each state $\boldsymbol \sigma_l$ is dependent on the measurement history $\outcomes^{(l)}$, the final state $\boldsymbol \sigma_k$ has a part that is independent of the measurement history, namely $f(\boldsymbol \sigma_0)$, and that can therefore be unentangled therefrom.

Now, consider our quantum algorithm for filtering.
The filtering computes the function 
$f(\mathcal D) 
= (\mathcal D', R, \predicate(\mathcal D))$,
where $\mathcal D' = \filter(\mathcal D)$ and $R = \mathcal D \setminus \mathcal D'$.
(See~\cref{eq:F,eq:filter-op}.)
Note that this is reversible, as supposed above (and we keep track of the removed edges $R_l$ for precisely this reason).
It consists of classical computations interspersed with quantum searches.
Suppose that each search were perfect, in the sense that measurement always yielded a marked state.
In that case, the circuit before measurement is exactly as in~\eqref{eq:U_M}, 
with uniform probability over the possible outcomes, so if we can do the search, then we can implement $U_{\outcomeSet_l}$.
More precisely, using fixed-point amplitude amplification (Thm. 27 of~\cite{gilyen2019quantum}) we can implement $U_{\outcomeSet_l}$ such that the final state is $\epsilon$-far from ideal in using $O(\log(1/\epsilon) / \delta)$ queries, where $\delta$ is the usual leading term of Grover search (e.g., $\sqrt{N}$).

If we want the final state of the whole quantum tree search to be within distance $O(1)$ from ideal and the filtering algorithm is run $T$ times with $g$ search each, then we need $\epsilon = O(T g)$.